\documentclass{emulateapj}

\slugcomment{ApJ accepted, 7 November 2005}
\shorttitle{Dust in Resonant Extrasolar Kuiper Belts}
\shortauthors{Wyatt}

\begin{document}

\title{Dust in Resonant Extrasolar Kuiper Belts
--- Grain Size and Wavelength Dependence of Disk Structure}

\author{M. C. Wyatt\altaffilmark{1}}
\affil{UK Astronomy Technology Centre, Royal Observatory,
       Edinburgh EH9 3HJ, UK}
\altaffiltext{1}{Current address:
  Institute of Astronomy, University of Cambridge,
  Madingley Road, Cambridge CB3 0HA, UK}
\email{wyatt@ast.cam.ac.uk}

\begin{abstract}
This paper considers the distribution of dust which originates in the
break-up of planetesimals that are trapped in resonance with a planet.
The effect of radiation pressure on small dust grains causes
their orbits and so their spatial distribution to be significantly
different to that of the parent planetesimals which previous work
has shown to be clumpy.
It is shown that there are three distinct grain populations:
\textbf{(I)} large grains ($>$ a few mm) have the same clumpy resonant
distribution as the planetesimals;
\textbf{(II)} moderate sized grains (a few $\mu$m to a few mm) are no
longer in resonance and have an axisymmetric distribution;
\textbf{(III)} small grains ($<$ a few $\mu$m) are blown out of the
system by radiation pressure immediately on creation and so have a density
distribution which falls off as $\tau \propto 1/r$, however the structure
of these grains can be further divided into two subclasses:
\textbf{(IIIa)} grains produced in the destruction of population
I grains that exhibit trailing spiral structure which emanates
from the resonant clumps;
and \textbf{(IIIb)} grains produced from population II
grains that have an axisymmetric distribution.
Since observations in different wavebands are sensitive to different
sized dust grains, multi-wavelength imaging of debris disks provides a
valuable observational test of models which explain structure seen in
sub-mm observations as due to resonant trapping of planetesimals.
For example, a disk with a collisional cascade dust size distribution
with no blow-out grains would appear clumpy in the sub-mm (which
samples population I grains), and smooth at mid- to far-IR
wavelengths (which sample population II grains).
The wavelength of transition from clumpy to smooth structure is
indicative of the mass of the perturbing planet.
The size distribution of Vega's disk is modeled in the light of
the recent Spitzer observations showing that collisions with the
large quantities of population III grains seen in the
mid- to far-IR may be responsible for the low levels of population
II grains in this system.
The origin of these population III grains must be in the destruction
of the grains seen in the sub-mm images, and so at high resolution and
sensitivity the far-IR and mid-IR structure of the Vega disk is predicted to
include spiral structure emanating from the sub-mm clumps.
Such structure could be detected with MIRI on the JWST and, if so, would
confirm the presence of a planet at 65 AU in the Vega disk as well
as determine the direction of its orbit.
\end{abstract}



\section{Introduction}
\label{s:intro}
Debris disks are the dust disks that are fed by the collisional grinding down
of extrasolar planetesimal belts that are analogous to the Kuiper Belt in
the Solar System (Backman \& Paresce 1993; Wyatt et al. 2003).
One of the most defining features of the debris disks that have been imaged
is that most of these disks are not smooth, but clumpy.
Such clumps are seen in sub-mm and mm images of the dust disks around Vega
(Holland et al. 1998; Koerner, Sargent \& Ostroff 2001; Wilner et al. 2002),
$\epsilon$ Eridani (Greaves et al. 1998; Greaves et al. 2005) and
Fomalhaut (Holland et al. 2003), as well as in optical and mid-IR images of
$\beta$ Pictoris (Kalas et al. 2000; Telesco et al. 2005).

The origin of these clumps has been widely debated.
One possible interpretation is that the clumps are in fact unrelated, and are
the chance superposition of background objects.
However, this has been ruled out on statistical grounds based on number counts   
of such objects (e.g., Holland et al. 1998; Telesco et al. 2005).
Another possibility that has been considered is that the clumps were created
relatively recently in collisions between large planetesimals;
features in the structure of the zodiacal cloud have been interpreted in this
way (Dermott et al. 2002; Nesvorn\'{y} et al. 2003).
However, the high mass of material seen in the clumps in disks imaged at long
wavelengths indicates that such an event would have to have involved two
planetesimals at least the size of Pluto, which given the expected frequency
of such events again rules out this interpretation on statistical grounds
(Wyatt \& Dent 2002; see also Kenyon \& Bromley 2005).
This mechanism remains a possibility for clumps seen at shorter wavelengths,
for which smaller colliding planetesimals are required to create detectable
clumps, for clumps seen at small orbital radii, and for those seen toward
relatively young systems like $\beta$ Pictoris (Telesco et al. 2005;
Kenyon \& Bromley 2005).

The favoured interpretation for the clumps seen in sub-mm and mm images
is that they are associated with material that is trapped in resonance with
a planet orbiting in the disk.
Two classes of model invoke planetary resonances to explain the clumps, and
they differ in the mechanism by which material ends up in the resonances.
In one model, dust migrates inward due to Poynting-Robertson (P-R)
drag and upon encountering a resonance, resonant forces temporarily halt the
migration causing a concentration of dust in the resonances.
This is the mechanism by which the Earth's clumpy resonant ring is thought
to have formed (Dermott et al. 1994).
The clumpy structures of the disks of Vega, $\epsilon$ Eridani and Fomalhaut,
have all been modeled in this way indicating the presence of planets more
massive than Saturn oribitng at several tens of AU from these stars
(Ozernoy et al. 2000; Wilner et al. 2002; Quillen \& Thorndike 2002;
Deller \& Maddison 2005).
However, these models suffer from the problem that, unlike dust in
the zodiacal cloud, inward migration due to P-R drag is not significant
in these systems because their disks are so dense that collisions occur
on much shorter timescales.
This means planetesimals are ground into dust fine enough to be removed
by radiation pressure before P-R drag has a chance to act (Wyatt 2005). 
This is not the case for dust coming from the asteroid belt which is
much less dense meaning that collisions are much less frequent.

In the other model, the parent planetesimals of the dust were trapped in
resonance with a planet which migrated outward early in the history of the
system, in the same way Kuiper Belt objects were trapped in resonance with
Neptune when it migrated outward (e.g. Malhotra 1995; Levison \& Morbidelli
2003).
Such migration could have been caused by angular momentum exchange when the
planetary system scattered the residual planetsimal disk (Fernandez \& Ip 1984;
Malhotra 1993; Hahn \& Malhotra 1999; Ida et al. 2000).
Wyatt (2003; W03) used numerical simulations to model the dynamical and spatial
structure of a planetesimal disk resulting from the outward migration of
a planet.
W03 also showed how comparing this structure with that seen in sub-mm images
of the Vega disk not only explained the observed structure without having to
invoke P-R drag, but also allowed constraints to be set on the planet causing
the structure, since only a specific range of planet mass and migration rate
can cause the observed structure.
A Neptune-mass planet which migrated from 40-65 AU over 56 Myr was proposed
which resulted in planetesimals being captured predominantly into the
planet's 2:1(u) and 3:2 resonances.
One of the limitations of the W03 model though was that the dust
seen in the sub-mm images was assumed to have the same distribution as
the planetesimals.
Small dust grains generally have different orbital parameters to
their parent planetesimals because radiation pressure causes them to
effectively see a lower mass star than did their parents.
This means that the dust distribution may be significantly different
to that of the planetesimals.

This paper considers how the distribution of dust arising from a
population of planetesimals previously trapped into resonance by a
migrating planet differs from that of the planetesimals themselves.
The distributions of two different types of grains are treated
separately: 
in \S \ref{s:bound}, numerical simulations are used to consider the
distribution of grains that remain gravitationally bound to the star;
\S \ref{s:unbound} models the collisional evolution of the disk to
determine the distribution of grains that are put on hyperbolic orbits
by radiation pressure as soon as they are created.
The findings are summarised in \S \ref{s:disc} which shows how the
disk can be divided into three grain populations sorted by grain
size, each of which exhibiting a distinct structure.
This section also considers which populations will dominate
observations in different wavebands and discusses the implications
for interpretation of the Vega disk in the light of the recent
Spitzer images of its structure (Su et al. 2005).
The conclusions are given in \S \ref{s:conc}.

\section{Distribution of Bound Grains}
\label{s:bound}
Small dust grains are acted on by a radiation force that can be
parametrised by a factor $\beta$ which is the ratio of the radiation
force to that of stellar gravity (Burns, Lamy, \& Soter 1979).
The factor $\beta$ is a function of particle diameter with smaller
particles having larger values of $\beta$ which for large grains
falls off $\propto 1/D$.
There are two components to the radiation force:
radiation pressure, which is the radial component that means that
the particle effectively sees a smaller mass star by a factor $1-\beta$;
and Poynting-Robertson (P-R) drag, which is the tangential component
that makes the particle's orbit spiral in toward the star at a rate
$\dot{a}_{pr} \propto -\beta/a$.
While it is included in the following numerical simulations, the
P-R drag force is not dominant in causing dust grains to
have different distributions to their parent planetesimals.
Rather that change in structure is caused by the fact that
on creation the dust grains have the same positions as their
parents, and similar velocities, but see a smaller mass star.
This causes the parameters describing their orbits to be different
to those of the parent planetesimal.
Dust grains created in the break-up of a planetesimal
that had $\beta = 0$, and for which orbital elements at the time of the
collision were $a,e,I,\Omega,\tilde{\omega}$, and $f$, move in the
same orbital plane as the parent, $I^{'} = I$ and
$\Omega^{'} = \Omega$, but on orbits with semimajor axes, $a^{'}$,
eccentricities, $e^{'}$, and pericenter orientations,
$\tilde{\omega}^{'}$, that are given by (Burns et al. 1979; Wyatt et al. 1999):
\begin{eqnarray}
  a^{'} & = & a(1-\beta) / \left[1 - 2\beta(1+e\cos{f})/(1-e^2)\right],
    \label{eq:aprime} \\
  e^{'} & = & (1-\beta)^{-1}\sqrt{e^2 + 2\beta e\cos{f} + \beta^2},
    \label{eq:eprime} \\
  \tilde{\omega}^{'} - \tilde{\omega} & = & f - f^{'} =
    \arctan{\left[ \beta \sin{f}/(\beta\cos{f} + e) \right]}.
    \label{eq:wprime}
\end{eqnarray}

Two types of dust grains are created:
those (large grains) with $\beta < 0.5$ that remain on bound orbits;
and those (small grains) with $\beta >0.5$ that are blown out of the
system on hyperbolic orbits as soon as they are created. 
For the former, the important question to ask is how a particle's
new orbital elements affect its subsequent evolution;
i.e., if the parent planetesimal was in a planet's resonance, is the
particle still in resonance, and if so how are the parameters describing
the resonant libration affected, which in turn tells us about the
spatial distribution of such dust grains (e.g., W03).
For the latter, the particles' subsequent evolution is not so important,
since these orbits are hyperbolic.
The more important issue is where those dust grains are most often created,
and so where their hyperbolic orbits start, and that is an issue which
depends on the collision rate of the parent planetesimals.
The former bound grains are considered in the remainder of this section,
while the latter hyperbolic (blow-out) grains are considered in
\S \ref{s:unbound}.

\subsection{Numerical Technique}
To derive the orbital parameters of small dust grains created in the
destruction of planetesimals previously trapped in resonance by
a migrating planet a three step process was used:
(i) first a population of parent planetesimals that were trapped in
resonance by a migrating planet was defined;
(ii) then the orbital elements of dust particles created in
collisions between those planetesimals was worked out;
(iii) then the dynamical evolution of those particles was followed
to quantify the effect of radiation pressure on their resonance
libration parameters.
This process was repeated until the libration parameters could
be determined for any size of dust grain associated with
planetesimals trapped in a planet's 3:2 and 2:1(u) resonances.
The new dust libration parameters were then used to work out the
spatial distribution of these grains.

The numerical technique employed in this paper is similar to that of
W03.
That is, numerical simulations were performed in which the dynamical
evolution of 200 massless objects and 1 planet of mass $M_{pl}$ was
followed using the RADAU fifteenth order integrator program (Everhart
1985).
All bodies are assumed to orbit a star of mass $M_\star = 2.5M_\odot$.
For some runs the adjustment described in W03 was employed that results
in a constant planet migration, $\dot{a}_{pl}$;
this adjustment was modified so that the migration rate decreased at
a linear rate to zero at the end of a migration.
The "objects" were either planetesimals or dust particles for which
their dynamical evolution is also affected by radiation pressure and
the P-R drag force characterised by the parameter $\beta$.

\subsubsection{Parent Planetesimal Distribution}
To define a population of planetesimals trapped in resonance, runs
were performed in which a planet was made to migrate through a disk
of 200 planetesimals at a rate starting at $\dot{a}_{pl}$ but decreasing
to 0 after a time $t_{mig}$.
At the start of the integration the planetesimals had eccentricities,
$e$, chosen randomly from the range 0 to $e_{max}=0.01$, and were
randomly distributed in semimajor axis between a narrow range
$a_1$ and $a_2$.
Their inclinations, $I$, were chosen randomly from the range 0 to
$e_{max}/2$ rad, and their arguments of periastron, $\tilde{\omega}$,
longitudes of ascending node, $\Omega$, and longitudes, $\lambda$, were each
chosen randomly from the range 0 to $360^\circ$.

The planet had all of these angles set to zero at the start of the
integration, with a semimajor axis $a_{pl1}$.
This meant that the planet ended up at a semimajor axis of
$a_{pl2} = a_{pl1}+\dot{a}_{pl}t_{mig}/2$.
The planet's migration rate was chosen so that all of the planetesimals
were trapped in the resonance being studied.
Trapping probabilities are a strong function of migration rate, with
slower migrations resulting in higher probabilities, and these
were derived in W03.
Since it was the strongest 3:2 and 2:1(u) resonances that were being
studied in this paper, the migration rate was chosen so that it was as
high as possible while still trapping all of the planetesimals into the
appropriate resonance;
slower migrations could have resulted in trapping into higher order
resonances.
The relationship between $a_{pl1}$ and $a_1$ was determined by
the requirement that the resonance being studied was in front of
the planetesimals at the start of the integration;
the location of the $p+q:p$ resonance is given by:
\begin{equation}
  a_r = a_{pl}[(p+q)/p]^{2/3}, \label{eq:ar}
\end{equation}
although the finite width of the resonance also had to be
taken into consideration.
The width of the initial planetesimal distribution, $a_2-a_1$, was
set to ensure that the planetesimals did not encounter the resonance
at the same phase, thus biasing the distribution of resonant
angles, $\phi$ (W03).
The extent of the migration is best described by its impact on
the mean semimajor axes and eccentricities of the planetesimal
population.
Higher eccentricities mean that more migration has taken place
by an amount that can be determined from the
eccentricity-semimajor axis relation
$e^2 = [q/(p+q)]ln(a/a_1)$ (W03).

\subsubsection{Initial Dust Distribution}
The initial dust distribution was taken directly from the orbits of
the planetesimal population at the end of the run.
In general each of these planetesimals would have a different
collision rate and so some of the 200 planetesimals could contribute
more to the overall dust population, however this effect was not
taken into account and 200 dust particles were produced with exactly
the same positions and velocities as the parent population.
Different sets of runs were then performed for dust particles
all of which have the same radiation pressure coefficient, $\beta$.
The initial orbital elements of the dust grains in each set of runs was
determined from equations (\ref{eq:aprime})-(\ref{eq:wprime}).
No further planet migration was assumed when considering the
evolution of the dust grains' orbits.

\subsubsection{Dust Orbital Evolution}
Even though the parent planetesimals were in resonance with $a=a_r$,
the dust particles may no longer be in exact resonance for two reasons.
First, the location of the resonance has changed because a dust
particle moves slower than a planetesimal at the same semimajor
axis because it sees a less massive star.
Thus to get the same ratio of orbital periods a dust particle
must be orbiting at:
\begin{equation} 
  a_{rd} = (1-\beta)^{1/3}a_r.
\end{equation}
Second, the semimajor axis of the particle has changed as described in
eq.~\ref{eq:aprime}.
Taking only terms to first order in eccentricity and $\beta$, it
is possible to show that the particle is at a semimajor
axis which is offset from the resonance by a factor
\begin{equation}
  \Delta a = a_d - a_{rd} \approx a_r \beta (4/3 \pm 2e),
  \label{eq:dapred}
\end{equation}
where the $\pm 2e$ term indicates whether the particle was created when
the planetesimal was at pericentre ($f=0$) or apocentre ($f=180^\circ$)
respectively.
In other words, smaller particles end up further from resonance by a factor
$\propto \beta$, and particles released at pericentre also end up further
from resonance than those released at apocentre.
This does not necessarily mean that the particles are no longer in
resonance, however, since resonances have finite width.

A particle is said to be in resonance if its resonant argument $\phi$ is
librating rather than circulating, where
\begin{equation}
  \phi = (p+q)\lambda_d - p\lambda_{pl} - q\tilde{\omega}_d,
\end{equation}
and libration can be characterised by a sinusoidal oscillation
\begin{equation}
  \phi = \phi_m + \Delta \phi \sin{2\pi t/t_\phi}
\end{equation}
of period $t_\phi$ and amplitude $\Delta \phi$, about a centre $\phi_m$.
Upon creation, the resonant argument of the dust particle is different to
that of the parent, since both the pericentres of the orbits, and the longitudes
of the particles within those orbits have changed (although in practise the
change in pericentre has the larger effect on the change in $\phi$).
Overall the change in $\phi$ is small unless a particle's $\beta$ is
close to, or larger than the eccentricity of the planetesimal $e$.
Since $e$ must be non-negligible to cause observable structure (W03),
and particles with even moderate values of $\beta$ are found to no longer
be in resonance (see later), this effect is not considered further.

It is also worth pointing out that the action of P-R drag is expected
to cause the resonant argument of the dust particle to librate about
a centre that is slightly offset from $180^\circ$ (or the appropriate
centre for the case of the 2:1 resonance, W03), since this is required for
the resonant forces to impart angular momentum to the particle to prevent
its inward orbital decay.
The new centre can be derived by making the sum of the inward decay
due to P-R drag (eq. 23 of Wyatt et al. 1999) and the semimajor
axis variation due to resonant forces (eq. 14 of W03) to zero, giving
\begin{equation}
  \phi_m - 180^\circ \propto \psi/\mu,
  \label{eq:phimpred}
\end{equation}
where $\psi = \beta \sqrt{M_\star/a}$ and $\mu = M_{pl}/M_\star$.
The parameter $\psi$ is equivalent to the parameter $\theta$ in
W03 in that it is the ratio of the dust particle's migration rate
to its orbital velocity, which determines the angle at which the
resonance is encountered.\footnote{To put the angle $\psi$
into perspective, a particle with $\beta=0.01$ at 1 AU from a
$1M_\odot$ star would meet the orbital velocity at an incident
angle of 0.4 arcsec (and would have $\psi=0.01$).}

To ascertain the impact of radiation pressure on the particles'
resonant arguments, the particles orbits were integrated for a
sufficient amount of time for several libration periods to be completed,
and so for the libration parameters to be fitted for each particle, and
the mean parameters for each population of 200 dust grains to be determined,
i.e., $\langle \phi_m \rangle$, $\langle \Delta \phi \rangle$,
and $\langle t_\phi \rangle$.
Again the libration periods of the parent planetesimals were
derived in W03, and while the periods of the dust grains do differ
from that of the parents, this provides a good enough idea to determine
the required integration times.
Since the oscillation also results in an oscillation of the particles'
semimajor axes, the amplitude of this oscillation was also determined,
$\Delta a$, as was the mean for the population of planetesimals,
$\langle \Delta a \rangle$.

\subsection{Results}

\begin{figure*}
  \centering
  \begin{tabular}{ccccc}
    & \hspace{-0.4in} $\beta = 0.002$ & 0.005 & 0.01 & 0.02 \\[-0.0in]
    \hspace{-0.25in} \includegraphics[width=1.32in]{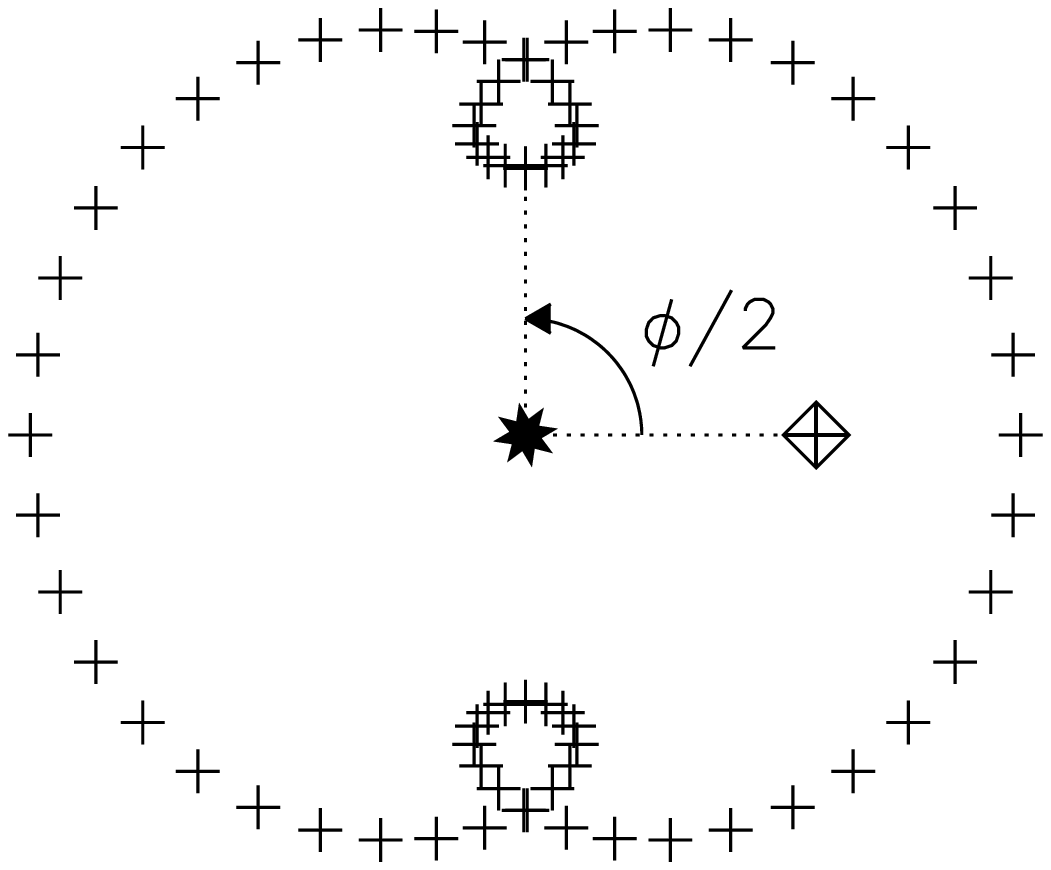} &
    \hspace{-0.5in} \includegraphics[width=1.32in]{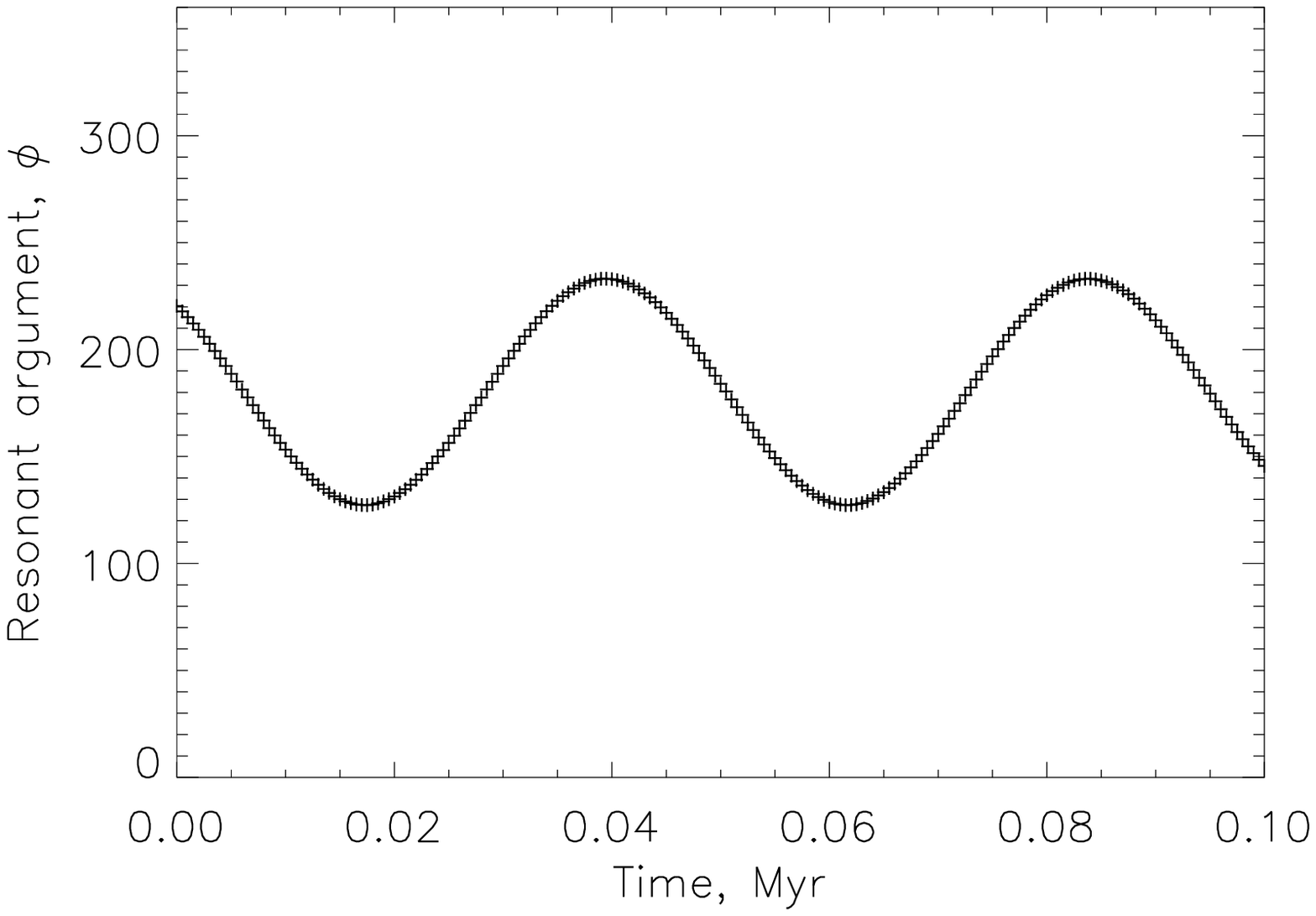} &
    \hspace{-0.05in} \includegraphics[width=1.32in]{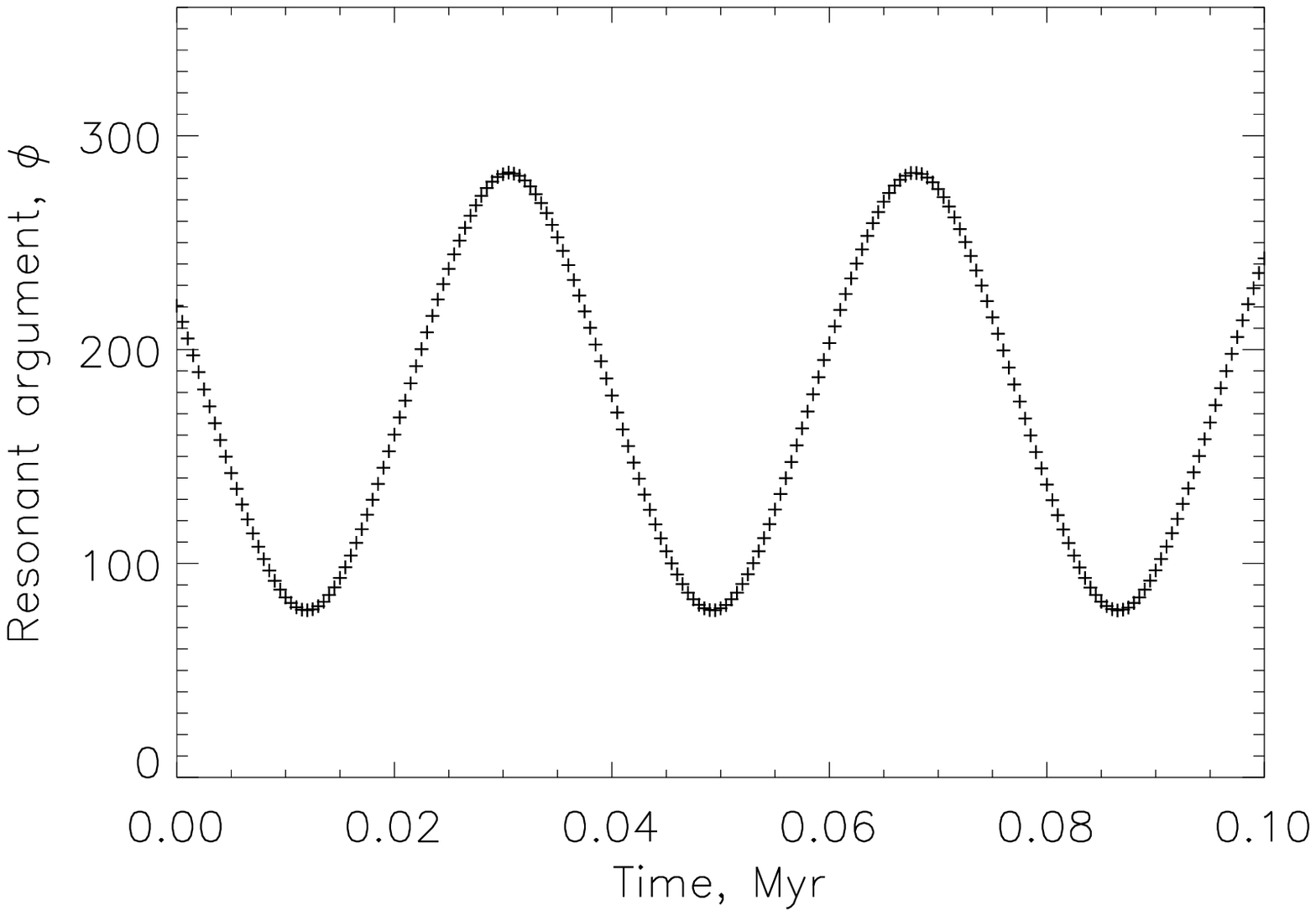} &
    \hspace{-0.05in} \includegraphics[width=1.32in]{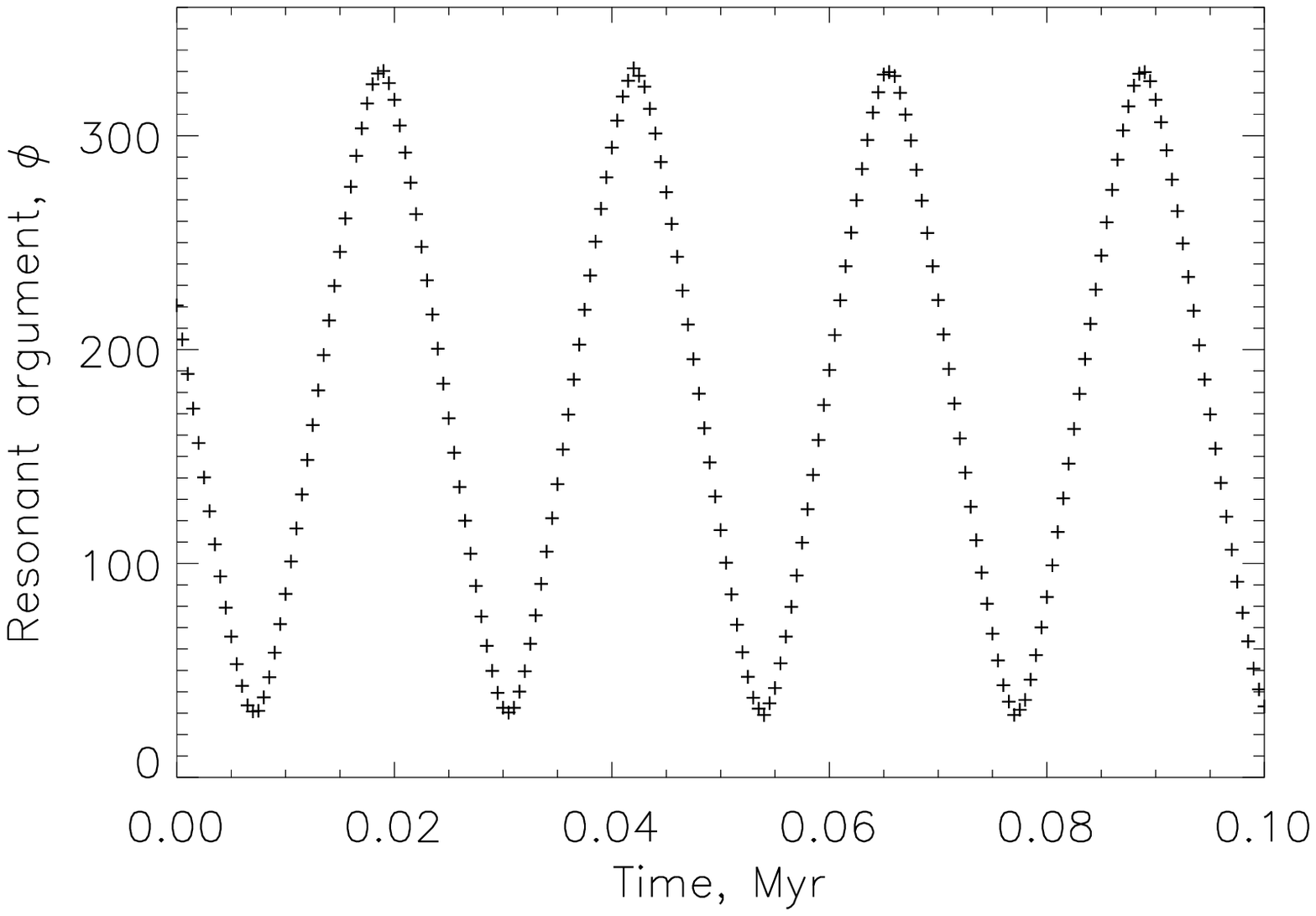} &
    \hspace{-0.05in} \includegraphics[width=1.32in]{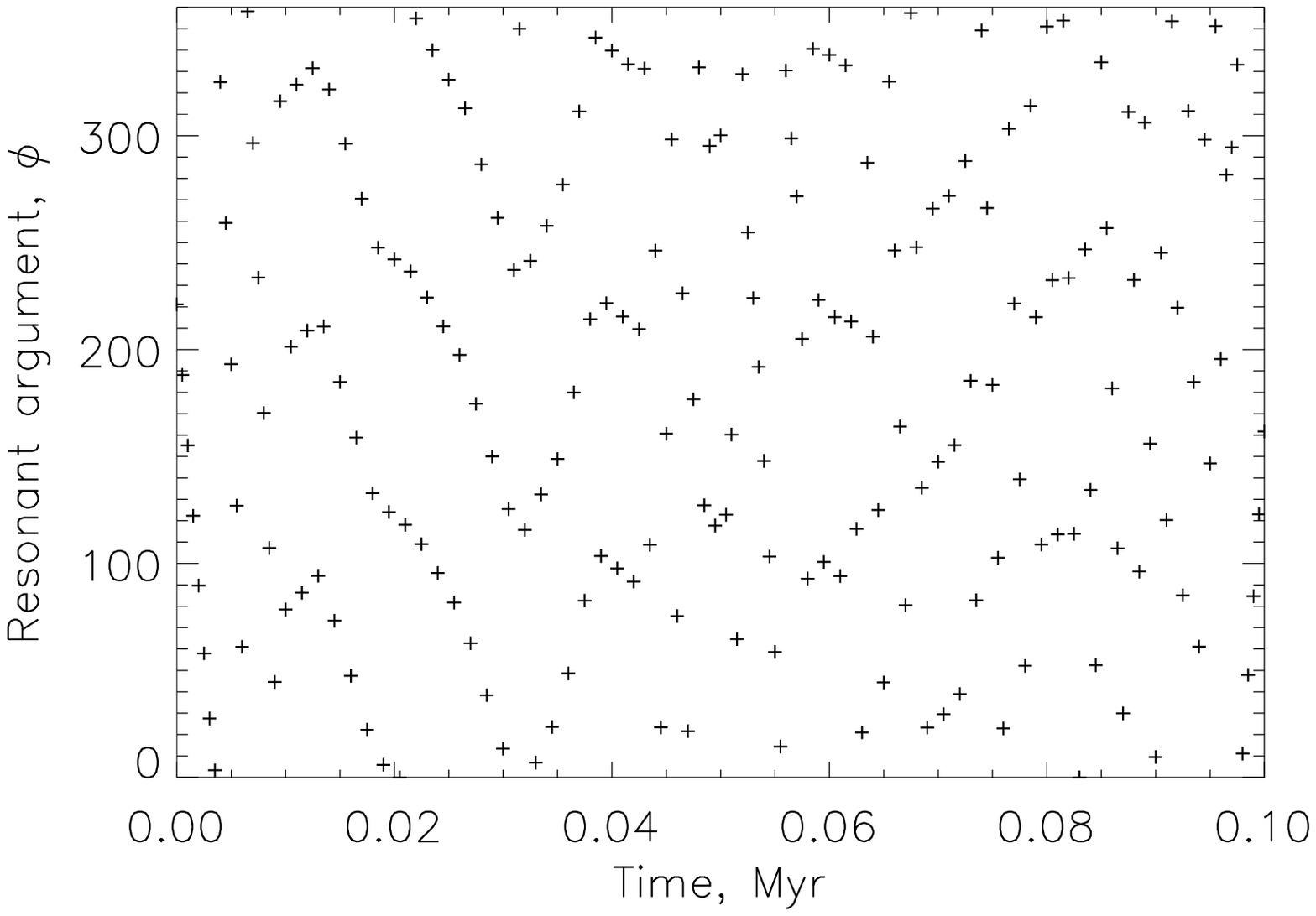} \\[-0.0in]
    \hspace{-0.05in} \includegraphics[width=1.32in]{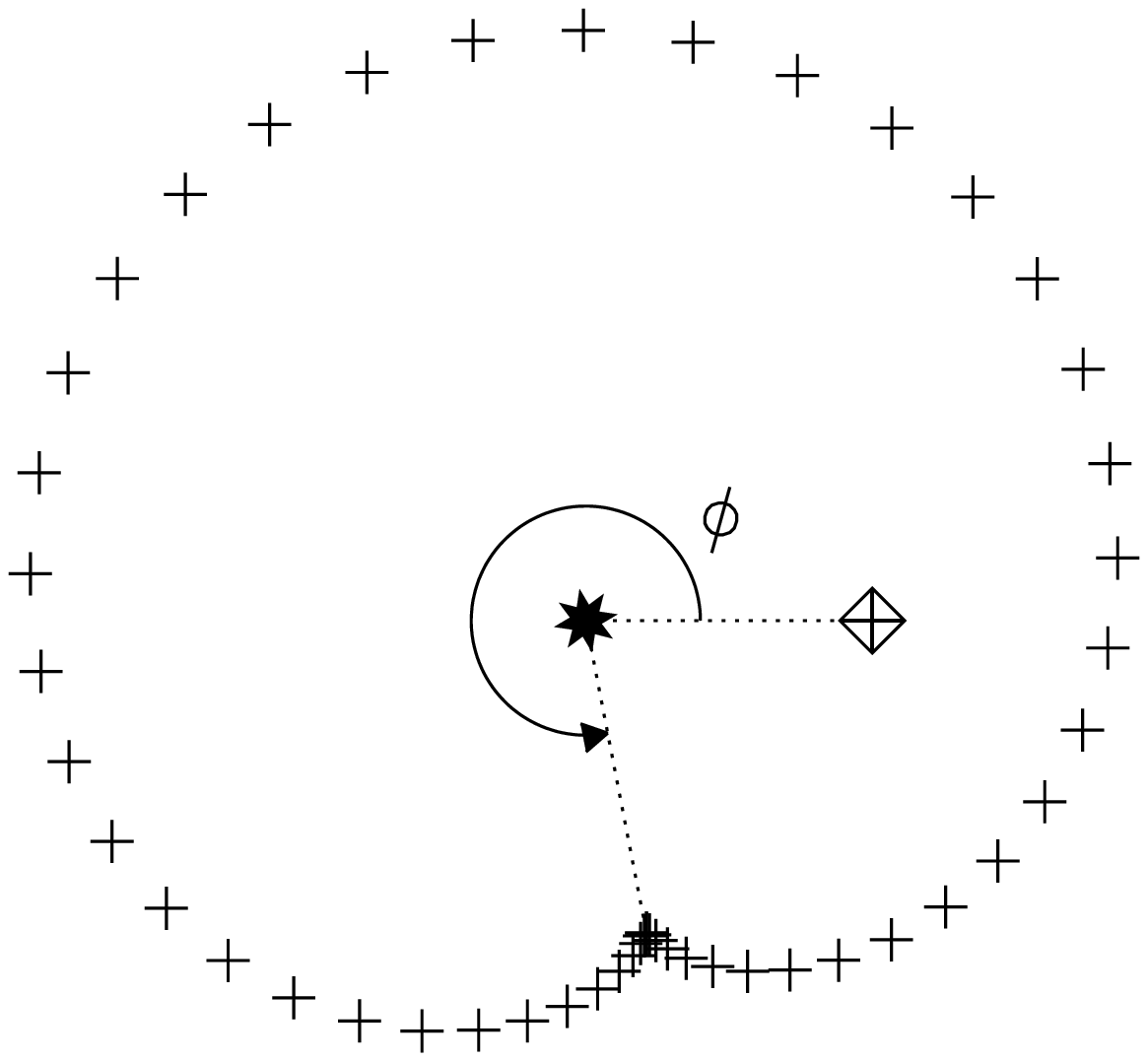} &
    \hspace{-0.5in} \includegraphics[width=1.32in]{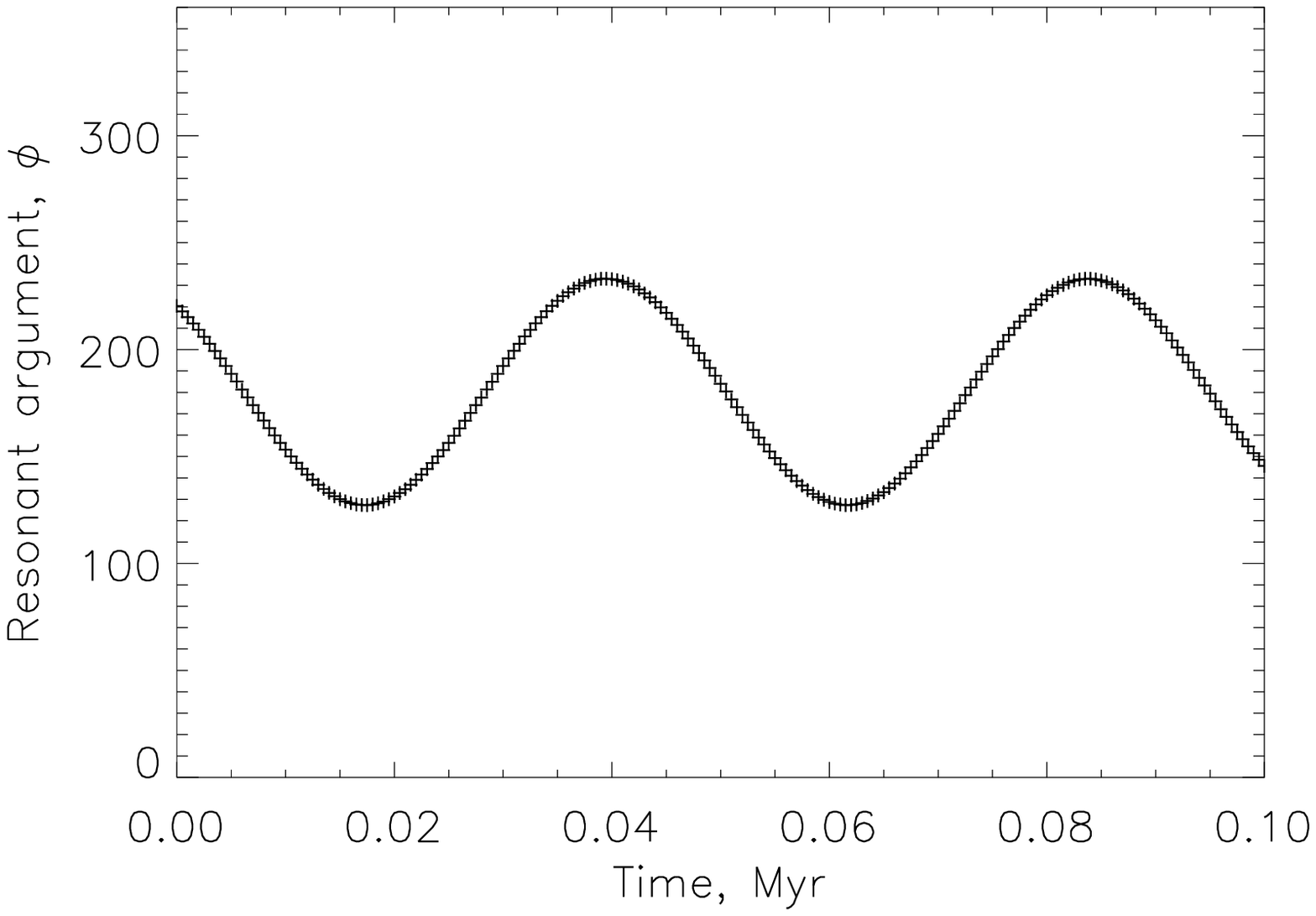} &
    \hspace{-0.05in} \includegraphics[width=1.32in]{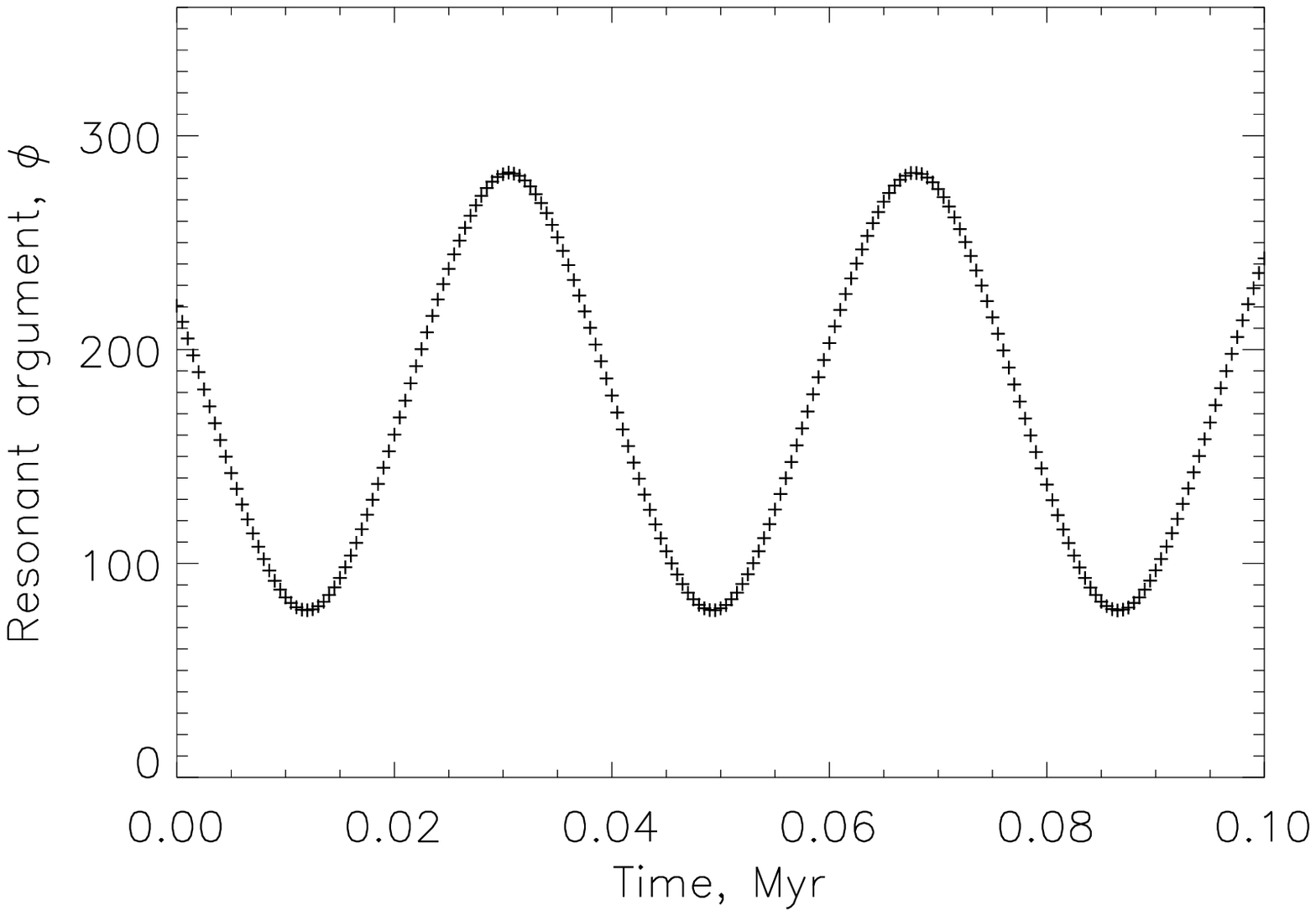} &
    \hspace{-0.05in} \includegraphics[width=1.32in]{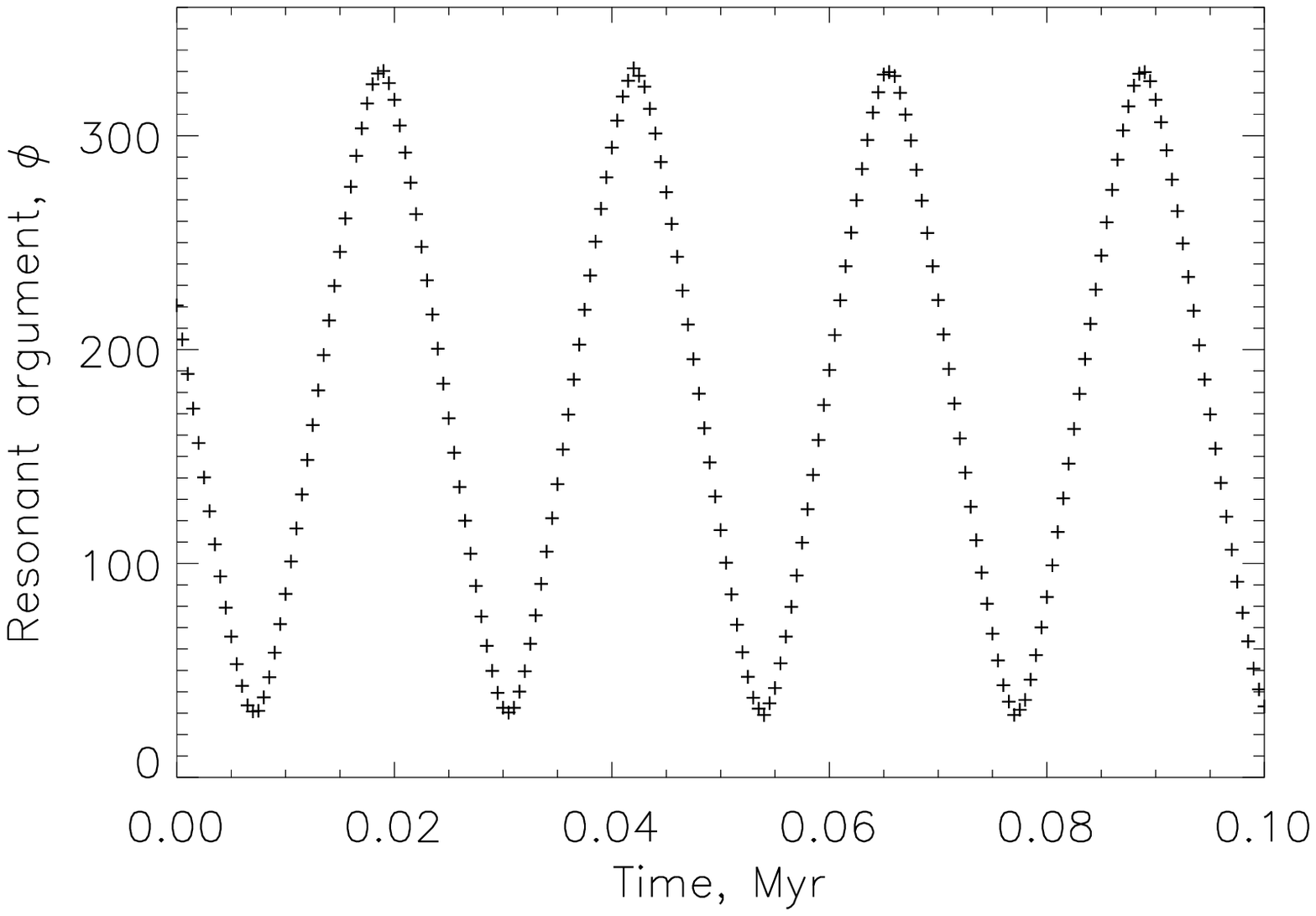} &
    \hspace{-0.05in} \includegraphics[width=1.32in]{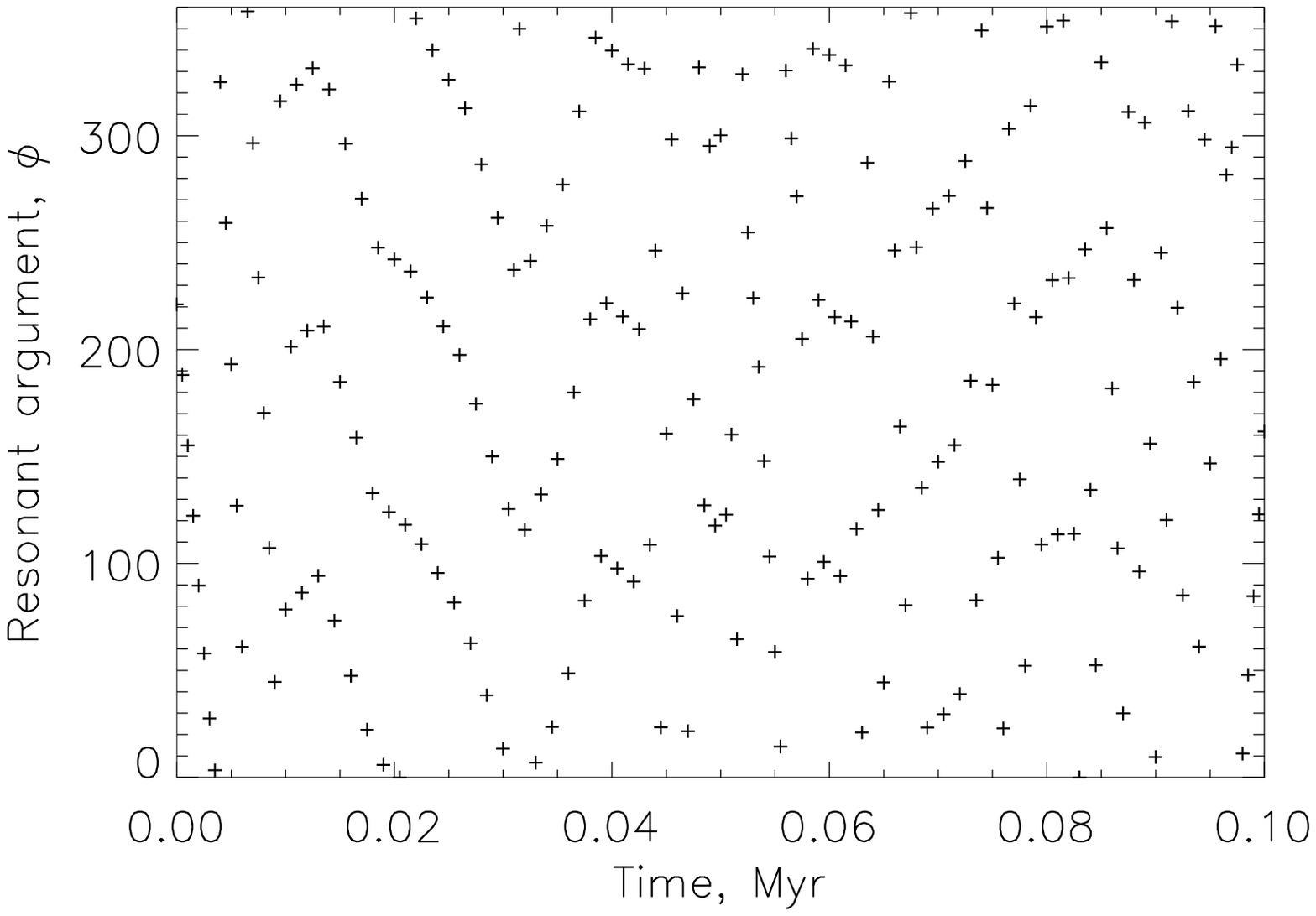}
  \end{tabular}
  \caption{Evolution of the resonant arguments, $\phi$, for different sized dust
  grains originating from the break-up of the same planetesimal in one of the runs for each
  resonance, and its implications for the spatial distribution of those grains.
  The top row is for the 3:2 resonance, for which dust grains exhibit a sinusoidal
  oscillation of $\phi$ with a libration width that increases for smaller grains
  until the particles fall out of resonance and $\phi$ circulates.
  The bottom row is for the 2:1(u) resonance, for which a similar progression is found, except
  that the libration is no longer sinusoidal (see text for discussion).
  The runs for both resonances correspond to parent planetesimals with eccentricities $\sim 0.3$
  that were trapped in resonance with a $30M_\oplus$ planet which migrated 45-60 AU from a
  $2.5M_\odot$ star.
  The far left plots show the path of resonant orbits in the frame corotating with then mean motion
  of the planet at equal timesteps for an orbit with an eccentricity of 0.3;
  the resonant arguments, $\phi$, determine the orientations of the loopy patterns of these orbits.}
  \label{fig:phievol}
\end{figure*}

\subsubsection{3:2 Resonance}

\begin{deluxetable}{ccccc}
  \tabletypesize{\scriptsize}
  \tablecaption{Parameters of the star ($M_\star$), planet ($M_{pl}$), parent
    planetesimals ($a$ and $e$), and dust particles ($\beta$) for runs
    characterising the libration parameters of dust grains created in the
    destruction of planetesimals previously trapped in the 3:2 resonance
    of a migrating planet.\label{tab:32runs}}
  \tablewidth{0pt}
  \tablehead{ \colhead{$M_\star, M_\odot$}  & \colhead{$M_{pl}, M_\oplus$}    &
    \colhead{$a$, AU}        &   \colhead{$e$}        &    \colhead{$\beta$} }
  \startdata
    2.5   & 10   &  78.6   &   0.28    &   0.0001 - 0.004  \\
    2.5   & 10   &  69.0   &   0.20    &   0.0001 - 0.004  \\
    2.5   & 10   &  64.4   &   0.13    &   0.0001 - 0.004  \\
    2.5   & 30   &  78.6   &   0.28    &   0.0001 - 0.005  \\
    2.5   & 100  &  78.6   &   0.28    &   0.0001 - 0.007  \\
    2.5   & 300  &  78.6   &   0.28    &   0.0001 - 0.01   \\
    0.5   & 10   &  78.6   &   0.28    &   0.0001 - 0.006  \\
    1.0   & 10   &  78.6   &   0.28    &   0.0001 - 0.004  \\
    1.5   & 10   &  78.6   &   0.28    &   0.0001 - 0.004  \\
    2.5   & 10   &  39.3   &   0.28    &   0.0001 - 0.006  \\
    2.5   & 10   & 196.5   &   0.28    &   0.0001 - 0.004 
  \enddata
\end{deluxetable}

The parameters used in the runs to characterise the libration parameters
of dust grains created in the destruction of planetesimals previously
trapped in the 3:2 resonance of a migrating planet are given in
Table \ref{tab:32runs}.
In this paper, a set of runs means those with the same star, planet and
planetesimals parameters, but different values of $\beta$ (corresponding
to different sized dust grains).
The first result is that for each set of runs, as $\beta$ is increased,
all of $\langle \phi_m \rangle$, $\langle \Delta \phi \rangle$,
and $\langle \Delta a \rangle$ also increase, while 
$\langle t_\phi \rangle$ decreases (see Fig.~\ref{fig:phievol}).
That is, for low $\beta$ (large grains) the dust particles remain in
resonance, but one with a libration centre that is further from $180^\circ$,
and with a higher libration width and range.
All of the increases are linear with $\beta$, although there is a turnover
in $\phi_m - 180^\circ$ for large $\beta$.
The runs found that particles remain in resonance
as long as $\Delta \phi < 180^\circ$.
Larger particles (those with higher $\beta$) are no longer in
resonance and for such grains $\phi$ circulates, i.e., undergoes
a monotonic decrease.

\begin{figure}
  \centering
  \begin{tabular}{cc}
    \textbf{(a)} & \hspace{-0.4in}
      \includegraphics[width=2.5in]{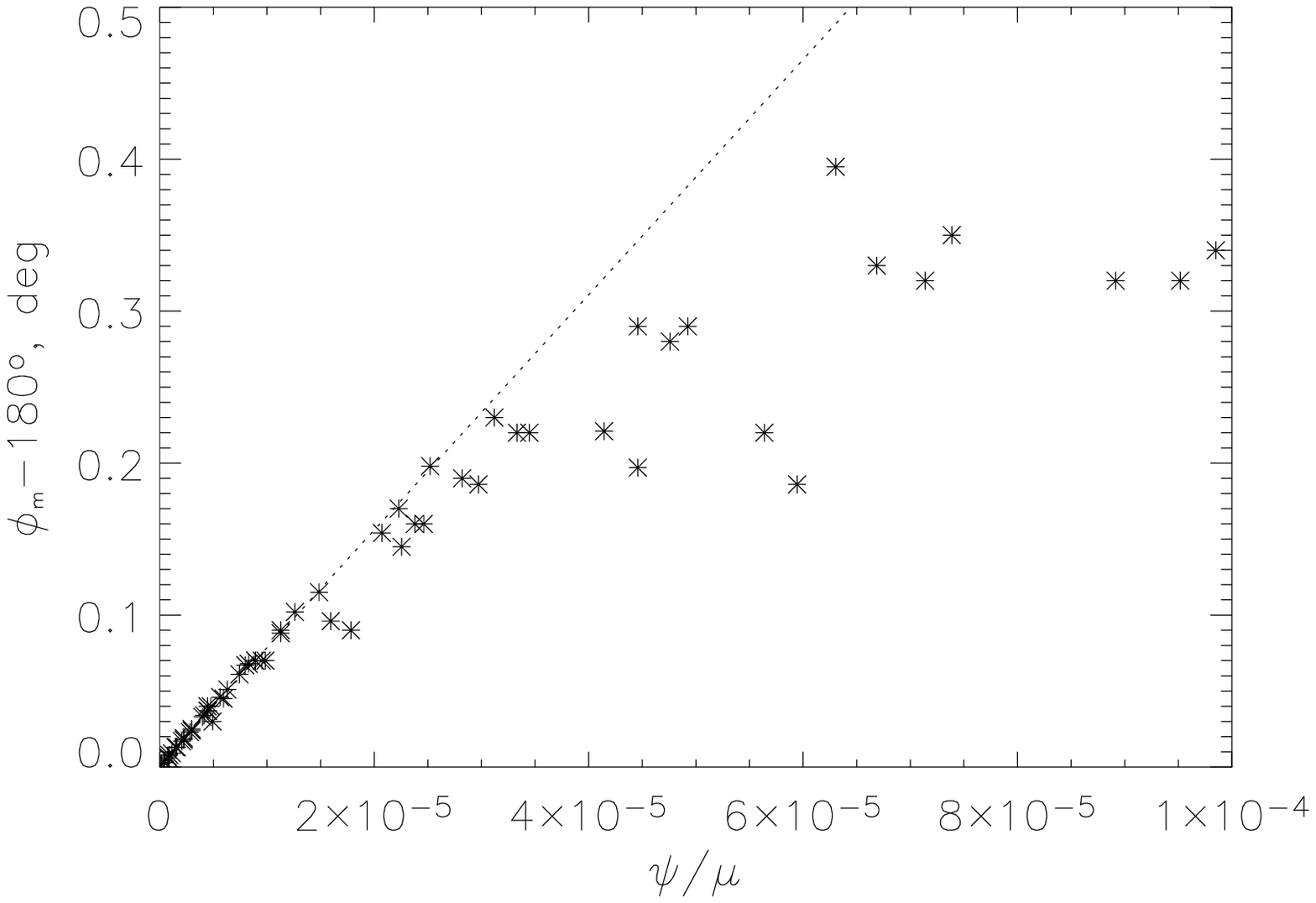} \\
    \textbf{(b)} & \hspace{-0.4in}
      \includegraphics[width=2.5in]{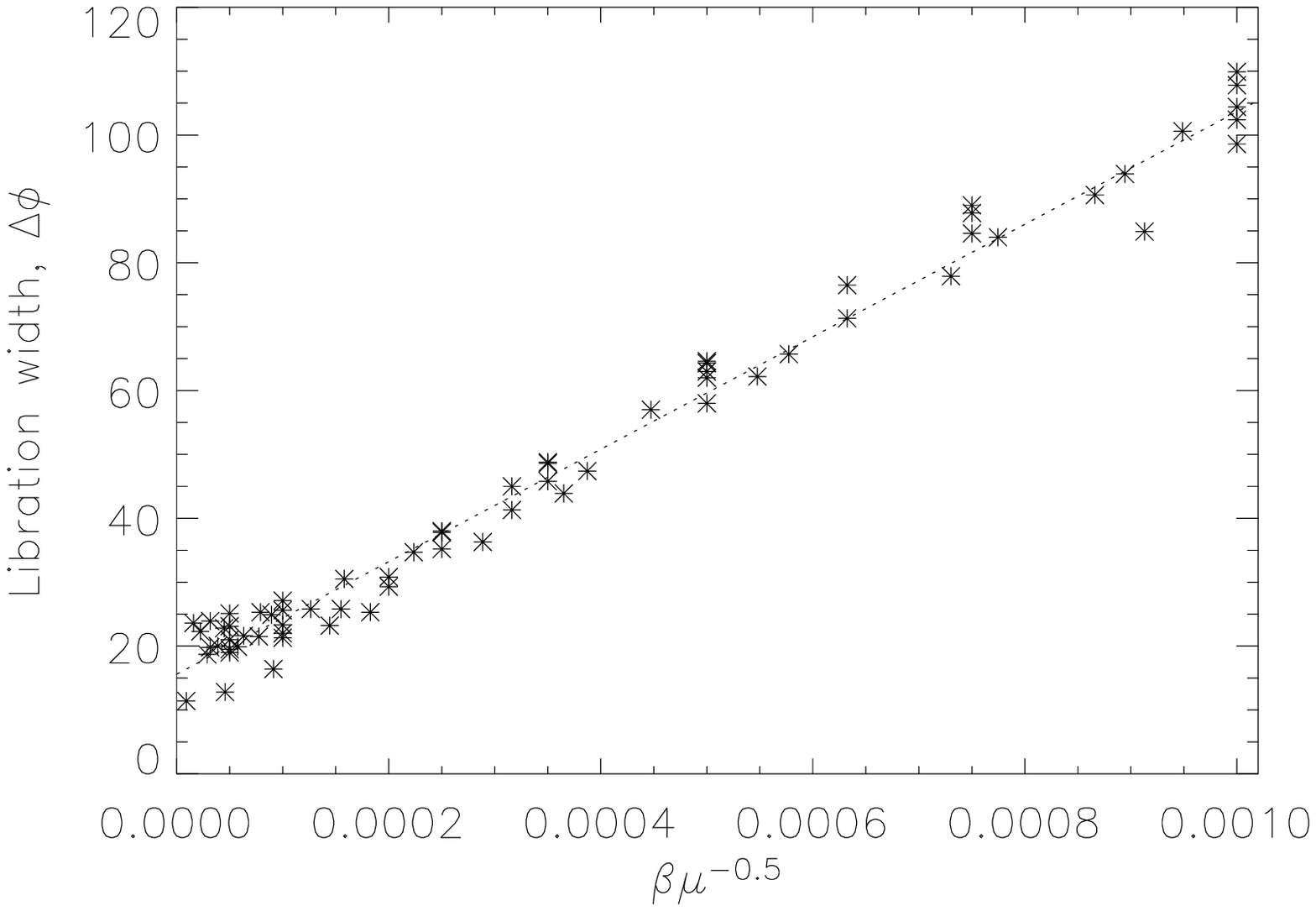} \\
    \textbf{(c)} & \hspace{-0.4in}
      \includegraphics[width=2.5in]{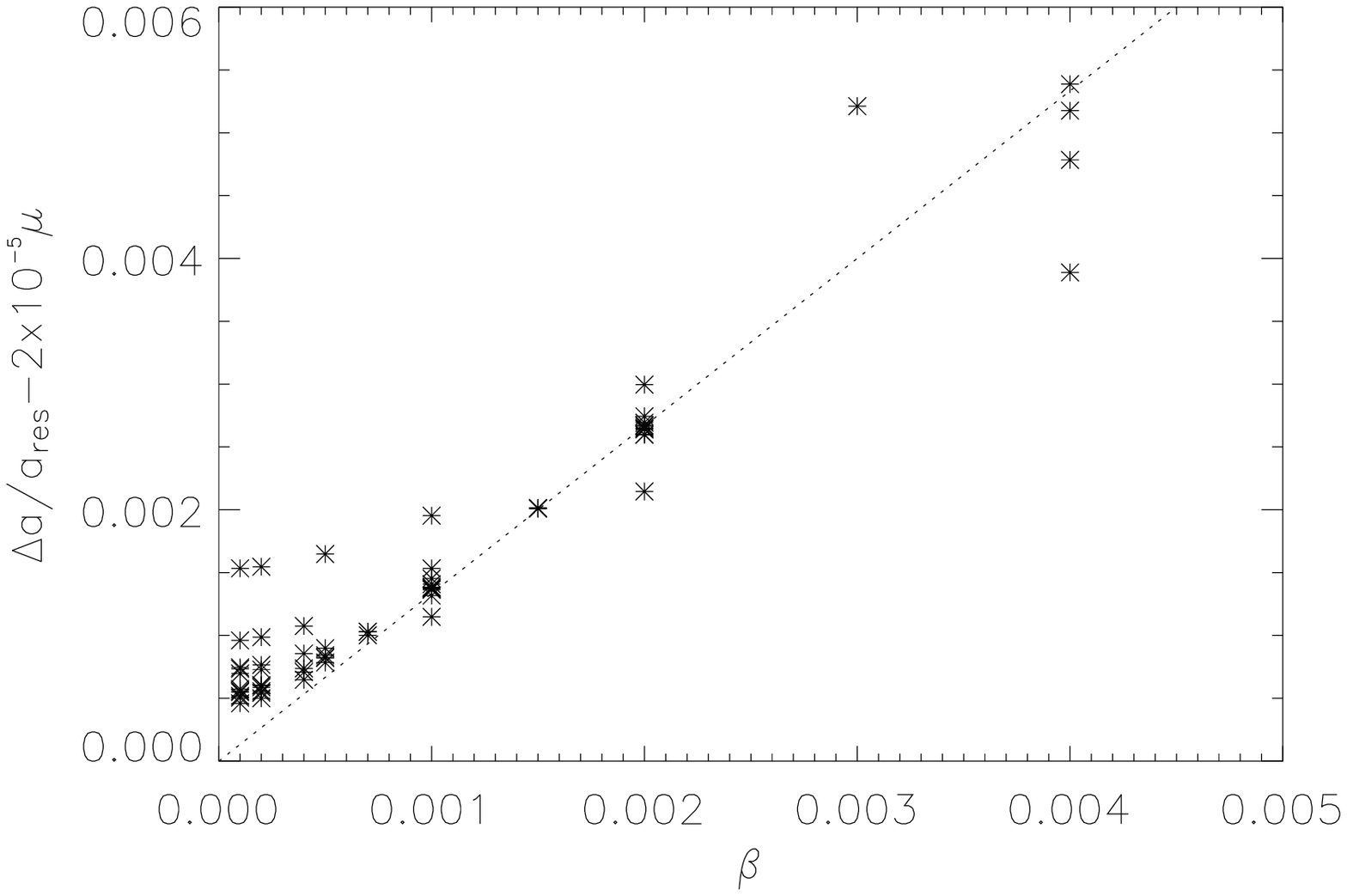}
  \end{tabular}
  \caption{Libration parameters for dust particles in the runs of Table
  \ref{tab:32runs}, i.e., for dust originating from planetesimals trapped
  in the 3:2 resonance of a migrating planet:
  \textbf{(a)} displacement of the libration centre, $\phi_m$, from $180^\circ$;
  \textbf{(b)} libration width, $\Delta \phi$;
  and \textbf{(c)} libration range, $\Delta a$.
  The fits from equations (\ref{eq:phimobs})-(\ref{eq:daobs})
  are shown with dotted lines.}
  \label{fig:32results}
\end{figure}

Different sets of runs were made varying each of the star, planet
and planetesimal parameters in turn to show the way the libration
of the dust grains was affected by each of these parameters (e.g., W03).
This showed that the displacement of the libration centre from $180^\circ$
depends on $\psi/\mu$ as predicted in equation (\ref{eq:phimpred}).
There is a turnover for high $\psi/\mu$, however the linear portion
of the curve can be well fitted by:
\begin{equation}
  \phi_m - 180^\circ = 7740 \psi/\mu,
  \label{eq:phimobs}
\end{equation}
where $\mu$ is in units of $M_\oplus/M_\odot$,
and this line is shown on Fig.~\ref{fig:32results}a along with the
results of all the runs.
Since the libration width is also high for high $\psi/\mu$, the
significance or meaning of the turnover is not clear, and may be an
artifact of the numerical method e.g., due to deviations from perfect
sinusoidal oscillation in this regime.

The increase in libration width could also be parametrised to explain
the results of all runs:
\begin{equation}
  \Delta \phi = 16^\circ + 88000\beta \mu^{-0.5},
  \label{eq:dphiobs}
\end{equation}
and this line is shown on Fig.~\ref{fig:32results}b along with the
results of all the runs.
In other words, the increased libration width is not dependent
on radial distance of the planet from the star, or on the star's
mass except in the ratio of the planet mass to the stellar mass, $\mu$.
The constant in this equation is indicative of the libration width
inherent in the planetesimal population (W03).
The increased libration range could also be explained by the
relation derived in equation (\ref{eq:dapred}) with the modification
that higher libration ranges result from different planet masses
according to:
\begin{equation}
  \Delta a / a_r = (4/3)\beta + 2 \times 10^{-5} \mu,
  \label{eq:daobs}
\end{equation}
and this result is represented in Fig.~\ref{fig:32results}c.
A libration range above that given in equation (\ref{eq:dapred}) is
expected if the particle's new semimajor axis is not the peak in 
its libration.
This is more likely to be the case for higher planet masses, since the
libration range is higher for such planets (Murray \& Dermott 1999).

The most important relationship in this paper is that given in
equation (\ref{eq:dphiobs}), since this can be used to
estimate the $\beta$ above which particles fall out of resonance,
which occurs when particles have $\Delta \phi > 180^\circ$.
In other words, grains are still in resonance as long as $\beta >
\beta_{crit}$, where:
\begin{equation}
  \beta_{crit} = 2 \times 10^{-3} \mu^{0.5}.
  \label{eq:betares}
\end{equation}
Since this corresponds to large grains, the black body approximation
can be used to estimate the grain size this corresponds to using
the relation $D = 0.4(L_\star/M_\star)/\beta$ in $\mu$m
(assuming a density of $\sim 2700$ kg m$^{-3}$; Wyatt et al. 1999) to
give:
\begin{equation}
  D_{crit} = 200(L_\star/M_\star)\mu^{-0.5}
  \label{eq:dres}
\end{equation}
in $\mu$m.
For example, dust grains arising from the destruction of the Plutino
population in the Kuiper Belt (objects in 3:2 resonance with Neptune)
remain in the resonance as long as they are larger than
$\sim 50$ $\mu$m.

\begin{figure*}
  \centering
  \begin{tabular}{c}
    \includegraphics[width=6.0in]{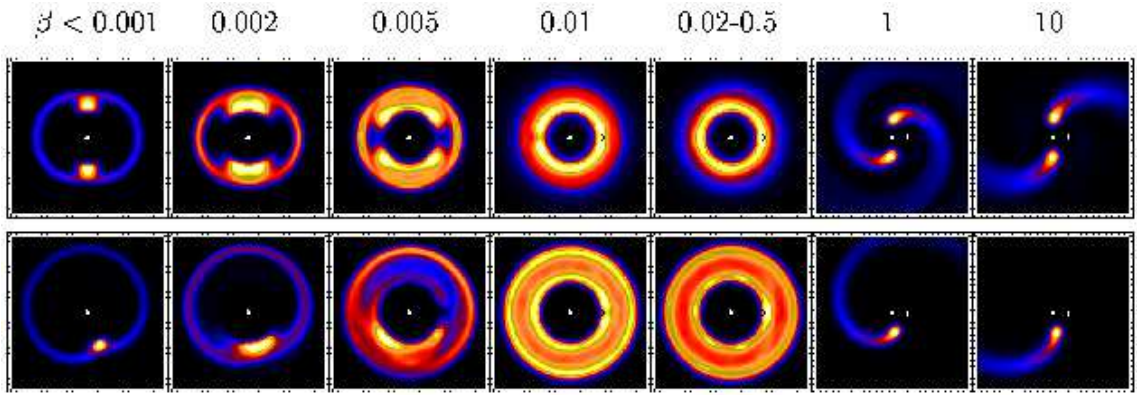}
  \end{tabular}
  \caption{Face-on view of the surface density distribution of dust grains of different sizes
  (characterised by the parameter $\beta$) created in the collisional destruction of planetesimals
  trapped in resonance with a $30M_\oplus$ planet which migrated
  45-60 AU from a $2.5M_\odot$ star:
  \textbf{(top)} dust from planetesimals in the 3:2 resonance with eccentricities of $\sim 0.28$;
  and \textbf{(bottom)} dust from planetesimals in the 2:1(u) resonance with eccentricities
  of $\sim 0.32$.
  The planet's location is shown with both a white plus and a black diamond, and its
  orbital motion is anticlockwise;
  the fields of view of the far right plots (for $\beta>0.5$) cover twice the area of the
  other ($\beta<0.5$) plots.
  The colour table is scaled linearly to the maximum density in the density in
  each plot.}
  \label{fig:spatial}
\end{figure*}

Equations (\ref{eq:phimobs}) and (\ref{eq:dphiobs}) also specify the
spatial distribution of dust grains of different sizes,
since Fig. 6 of W03, a modified version of which is reproduced here
in Fig.~\ref{fig:phievol}, showed how the angle $\phi$ determines
the orientation of the loopy pattern that a resonant object's orbit
makes in the frame rotating with the planet.
For this paper the only important parameter describing the
distribution of the resonant arguments of the dust grains is
the increased libration width, since one implication of equation
(\ref{eq:betares}) is that the offset of the libration centre from
$180^\circ$ is always small:
even the smallest grains still in resonance have
$\phi_m - 180^\circ < 15^\circ \sqrt{M_\star/a}\mu^{-1.5}$, so
this can be ignored unless the planet is of very low mass and
orbiting close to the star (such as dust grains trapped in
resonance with the Earth, Dermott et al. 1994).
The increased libration width of the orbits of resonant dust
grains causes the clumps they make to be azimuthally smeared
out.
This is illustrated in Fig.~\ref{fig:spatial} which shows the
distributions in the frame rotating with the planet of dust grains of
different sizes in one set of runs.
\footnote{To make these figures, the locations of the dust grains
were coadded at all times in the frame rotating with the planet,
and the integration was long enough to cover a large number of
libration periods so that the structure was independent of the
length of the integration (100,000 year integrations are shown
in Fig.~\ref{fig:spatial}).}

Fig.~\ref{fig:spatial} also illustrates how particles that are
no longer in resonance have an axisymmetric distribution, and
as expected this occurs for particles with $\beta > \beta_{crit}$
(eq.~\ref{eq:betares}).
While the particles are not in resonance, the resonant argument
is still relevant to this discussion, since the pattern of the
particle's orbit in the frame rotating with the planet is still
close to that shown in Fig.~\ref{fig:phievol}.
The particles' new semimajor axes mean that there is a
monotonic decrease in the value of $\phi$, which if
additional perturbations due to resonant forces are ignored
can be estimated to be:
\begin{equation}
  \dot{\phi}_d = -720^\circ p\beta(1 \pm 1.5e)/t_{pl},
  \label{eq:phiddot}
\end{equation}
where $p=2$ for the 3:2 resonance.
This means that the loopy pattern rotates so that the
the clump in front of the planet's motion approaches the
planet (with the other clump $180^\circ$ away).
The timescale on which a non-resonant particle has its pericentre at the
same longitude as the planet can be estimated from equation
(\ref{eq:phiddot}), since for this to happen $\phi$ would have
to decrease from $180^\circ$ to $0^\circ$ which occurs on a
timescale of
\begin{equation}
  t_{scat} = 180^\circ/|\dot{\phi}_d| = 0.25t_{pl}/[p\beta(1 \pm 1.5e)].
  \label{eq:tscat}
\end{equation}

This is an important timescale because if the particle's
eccentricity is high enough, a close approach to the planet is possible
causing the particle to be scattered onto a more eccentric and inclined
orbit.
\footnote{Note that in the absence of close approaches, such as might
be the case for dust coming from planetesimals with low eccentricity,
it can be shown that P-R drag would cause particles to reattain the
resonance of their parent planetesimals on a timescale that is
independent of particle size: $t_{pr} = 1070 a_r^2/M_\star$ in years.}
Since after scattering the particle's pericentre (or apocentre) remains
close to the orbit of the planet further close approaches and scattering
can then ensue.
This scattering process was studied by following the simulations of
$\beta=0.02$ and 0.2 grains coming from the planetesimals in the
3:2 resonance shown in Fig.~\ref{fig:spatial} for 350 Myr, though
excluding grains once their semimajor axis dropped below 40 AU
(either because such grains would be scattered by interior planets
or evolve due to P-R drag on to the star).
These simulations showed that some grains were inserted quickly into
the resonances outside the 3:2 resonance (some $\beta=0.02$ grains
were trapped in the 8:5 resonance, while the $\beta=0.2$ grains
populated a large number of resonances including the 2:1, 5:2, 3:1,
7:2, 4:1), in which case they remained there until P-R drag
forces had increased their eccentricities to the maximum value
for the resonance at which point
the libration width increased until the resonance was unstable,
a process which took $\sim 30-50$ Myr.
The non-resonant grains were scattered relatively quickly (on timescales
as low as $t_{scat}$), but remained in the system to undergo further
scattering events.
These grains were eventually excluded following scattering events
which put their semimajor axes below that of the planet, and once
P-R drag had further reduced their semimajor axes below 40 AU, a
process which took a total of $\sim 10-30$ Myr.

The resulting spatial distribution of the grains is not only
axisymmetric, but also more radially and vertically extended than
that of the parent planetesimals.
The resonances did introduce a small clumpiness in the resulting
distribution, however due to the variety of resonances populated
by a small fraction of particles, this is expected to have a
minimal effect on the distribution, resulting in only a
slight underdensity at the location of the planet for these grains.
The scattering timescale itself (eq.~\ref{eq:tscat}) is relatively
short, since grains with $\beta=0.5$ to $\beta_{crit}$ encounter
the planet on timescales of $t_{scat} = 0.25t_{pl}$ to
$62.5\mu^{-0.5}t_{pl}$.
While this is likely to be significantly shorter than the collisional
timescale, the simulations showed that the dynamical lifetime of the
grains may be much longer than their collisional lifetime.
The scattering timescale given in eq.~\ref{eq:tscat} should thus be
considered as the timescale on which the particle distribution becomes
axisymmetric.

\subsubsection{2:1 Resonance}
\label{sss:21}

\begin{deluxetable}{ccccc}
  \tabletypesize{\scriptsize}
  \tablecaption{Parameters of the star ($M_\star$), planet ($M_{pl}$), parent
    planetesimals ($a$ and $e$), and dust particles ($\beta$) for runs
    characterising the libration parameters of dust grains created in the
    destruction of planetesimals previously trapped in the 2:1(u) resonance
    of a migrating planet.\label{tab:21uruns}}
  \tablewidth{0pt}
  \tablehead{ \colhead{$M_\star, M_\odot$}  & \colhead{$M_{pl}, M_\oplus$}    &
    \colhead{$a$, AU}        &   \colhead{$e$}        &    \colhead{$\beta$} }
  \startdata
    0.8   & 10   & 79.4   &   0.44    &   0.0001 - 0.1  \\
    0.8   & 30   & 79.4   &   0.44    &   0.0001 - 0.1  \\
    0.8   & 100  & 79.4   &   0.44    &   0.0001 - 0.1  \\
    0.8   & 300  & 79.4   &   0.44    &   0.0001 - 0.1  \\
    2.5   & 30   & 95.3   &   0.32    &   0.005 - 0.02  \\
    2.5   & 300  & 95.3   &   0.32    &   0.01 - 0.03  \\
    2.5   & 300  & 31.8   &   0.32    &   0.01 - 0.03  \\
    0.25  & 30   & 95.3   &   0.32    &   0.01 - 0.03  \\
    2.5   & 300  & 95.3   &   0.43    &   0.01 - 0.03  \\
    2.5   & 300  & 95.3   &   0.20    &   0.01 - 0.03  \\
    2.5   & 300  & 95.3   &   0.12    &   0.005 - 0.03
  \enddata
\end{deluxetable}

The parameters used in the runs to characterise the libration parameters
of dust grains created in the destruction of planetesimals previously
trapped in the 2:1 resonance of a migrating planet are given in
Table \ref{tab:21uruns}.
For these runs, the migration was set so that planetesimals were trapped
only into the 2:1(u) resonance (i.e., with $\phi$ librating about $\sim
270^\circ$ causing a clump of material $\sim 90^\circ$ behind the planet's
motion).
The results of the runs for the 2:1 resonance are broadly similar to those
for the 3:2 resonance in that for increasing $\beta$
(decreasing grain size), the libration width increases leading to
clump smearing and eventually an axisymmetric spatial distribution.

However, one significant difference was that as the particles $\beta$
was increased, the libration of $\phi$ could no longer be described as a
sinusoidal oscillation.
This is illustrated in Fig.~\ref{fig:phievol} which shows the
evolution of $\phi$ for different sized dust grains coming from the
same planetesimal in one of the runs, as well as Fig.~\ref{fig:spatial}
which shows the spatial distributions of different sized dust grains
coming from the same poulation of planetesimals.
As $\beta$ is increased, first the libration becomes asymmetric in that
more time is spent at low values of $\phi$;
this means that the clump smears out asymmetrically about $\sim
270^\circ$, causing a spatial distribution weighted toward the
anti-planet direction.
Increasing $\beta$ further leads to the particle switching between
performing half a libration in each of the 2:1(u) and 2:1(l) resonances.
The distribution of such grains has three defining features:
a gap at the location of the planet;
concentrations just in front of and behind the planet (from the
extreme points of the libration);
and an overdensity in the anti-planet direction, caused by resonant
forces which slow down the evolution of $\phi$ at $\sim 180^\circ$
(this is just noticeable in the $\beta=0.01$ run of
Fig.~\ref{fig:phievol}, but dominates the evolution of some grains).
Finally the particle is no longer in resonance at which point $\phi$
circulates rather than librates resulting in an axisymmetric distribution.

The non-sinusoidal oscillation makes characterisation of the structure
more problematic than for the 3:2 resonance, and for the purposes of
this paper the runs were used to derive the $\beta$ for which the
libration width is so high that grains with $\beta>\beta_{crit}$ are no
longer in resonance.
The result is that the stability of grains from planetesimals in the
2:1(u) resonance is remarkably similar to those from the 3:2 resonance.
In fact, $\beta_{crit}$ can be well approximated by equation
(\ref{eq:betares}) as illustrated in Fig.~\ref{fig:spatial}, since the
cut-off has the same dependence on $M_{pl}$, $M_\star$ and $a$, and
occurs at the same size of dust grain.
One difference with the 2:1(u) resonance is that there is a dependence on
particle eccentricity, in that $\beta_{crit} \propto e^{1.0\pm0.5}$;
i.e., dust from higher eccentricity planetesimals remains in resonance down
to smaller grains.
Equation (\ref{eq:betares}) is thus valid for the 2:1(u) resonance when
the eccentricities are $\sim 0.3$.
However, since high eccentricities are required to cause significant
structure (W03), the difference caused by this effect is likely to be
small and equation (\ref{eq:betares}) is used in this paper as a good
approximation for the cut-off.

\section{Distribution of Blow-out Grains}
\label{s:unbound}

\begin{figure}
  \centering
  \begin{tabular}{cc}
    \hspace{-0.1in} \includegraphics[width=1.8in]{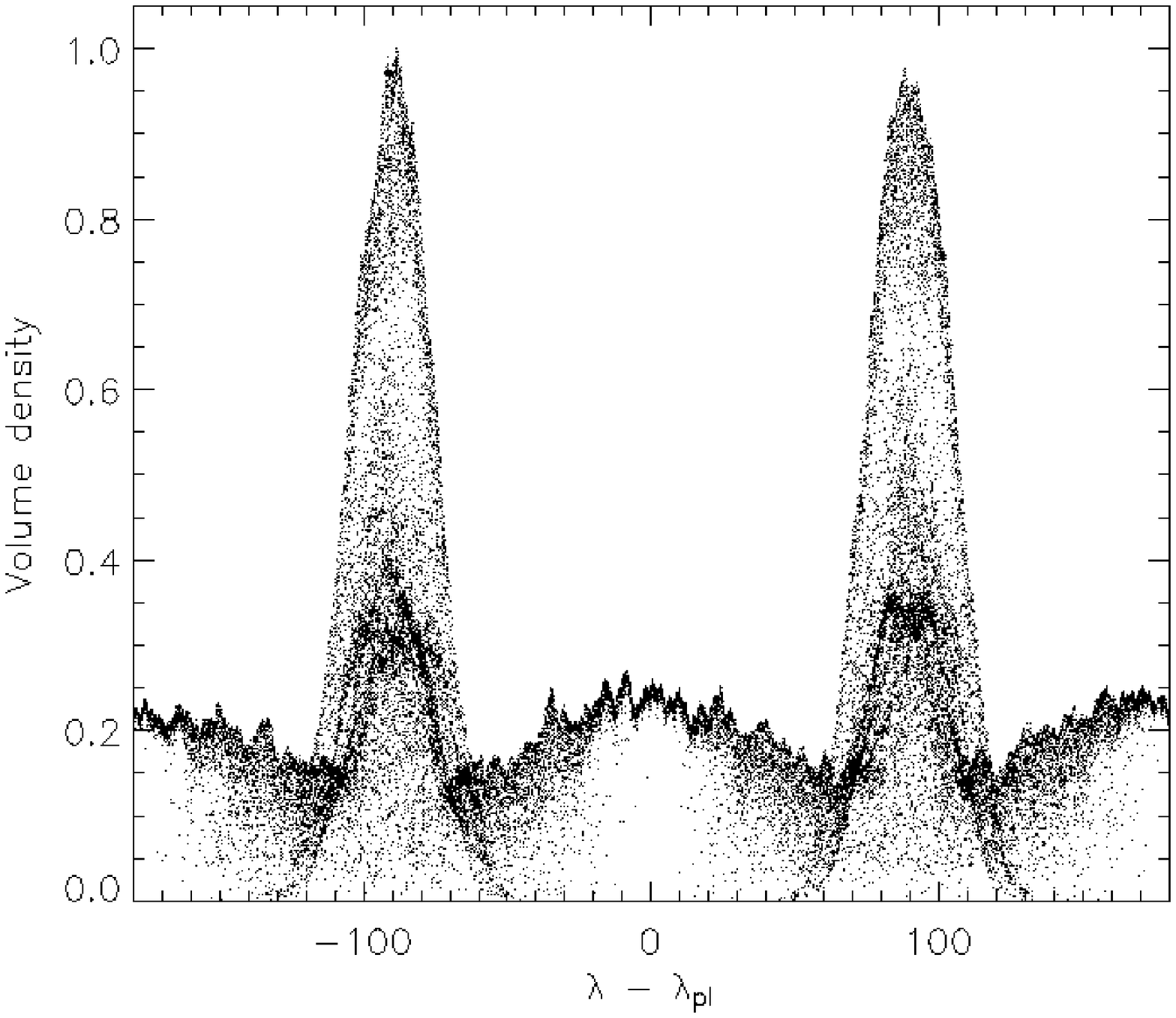} &
    \hspace{-0.2in} \includegraphics[width=1.8in]{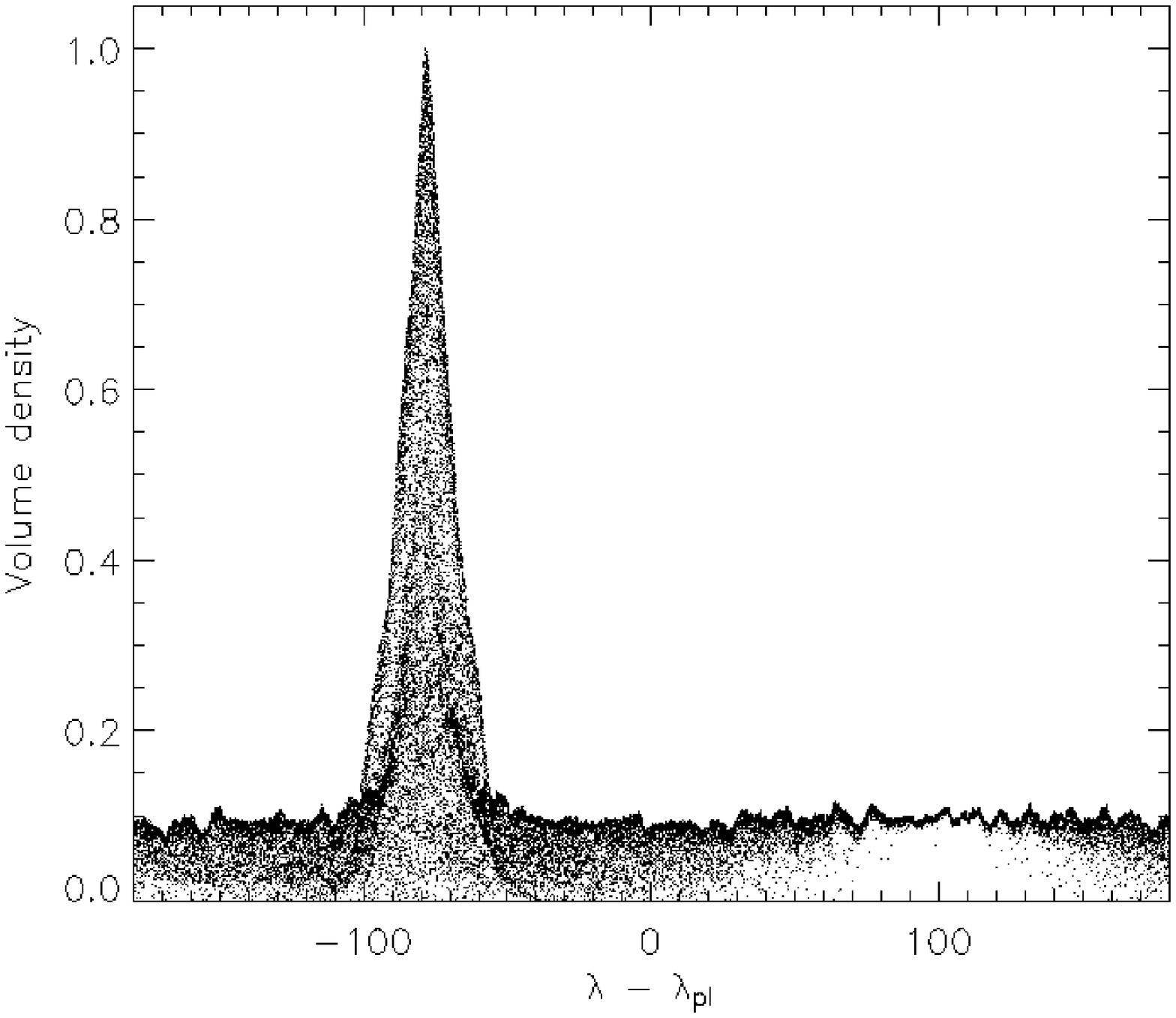} \\[-0.1in]
    \hspace{-0.1in} \includegraphics[width=1.8in]{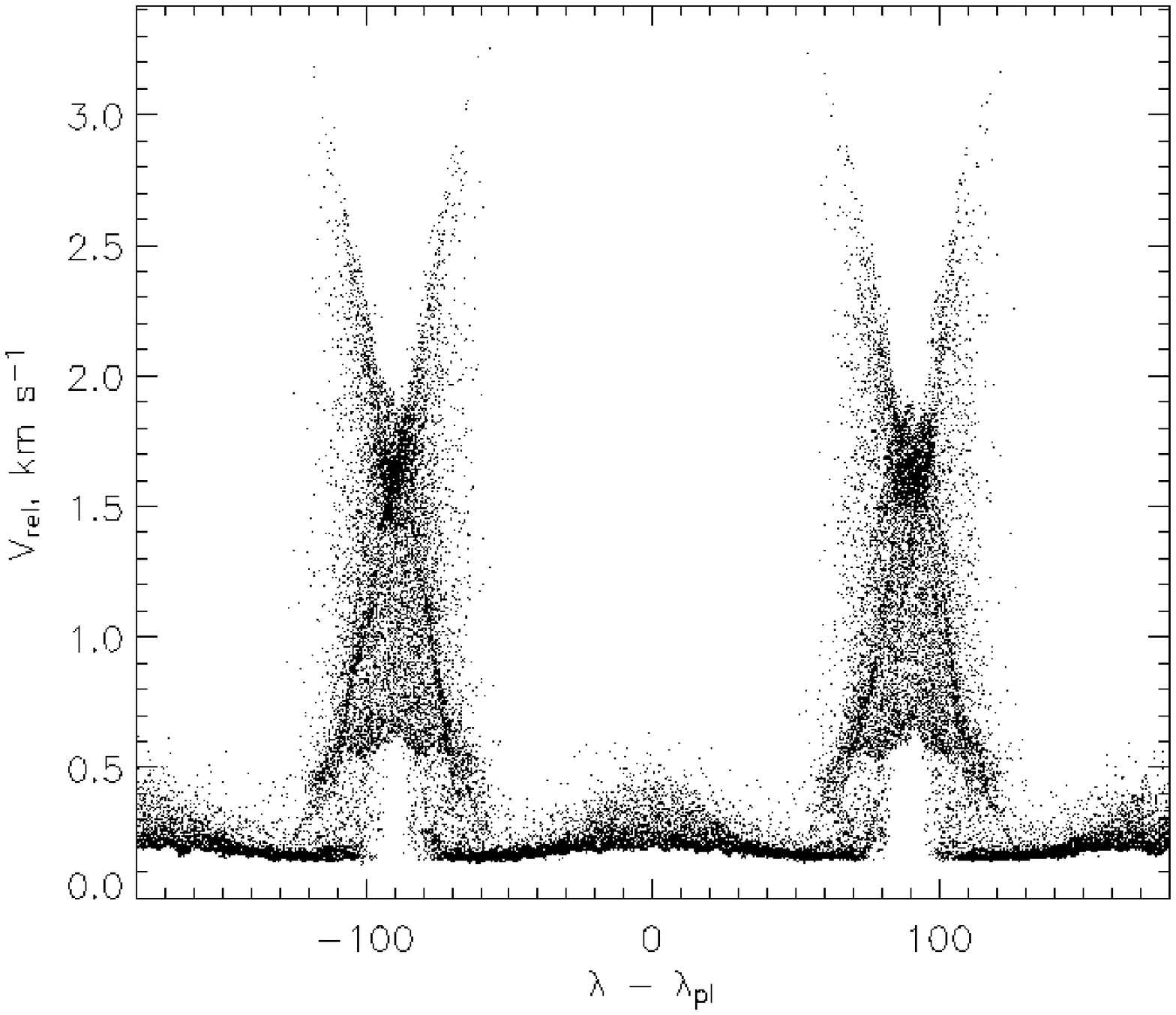} &
    \hspace{-0.2in} \includegraphics[width=1.8in]{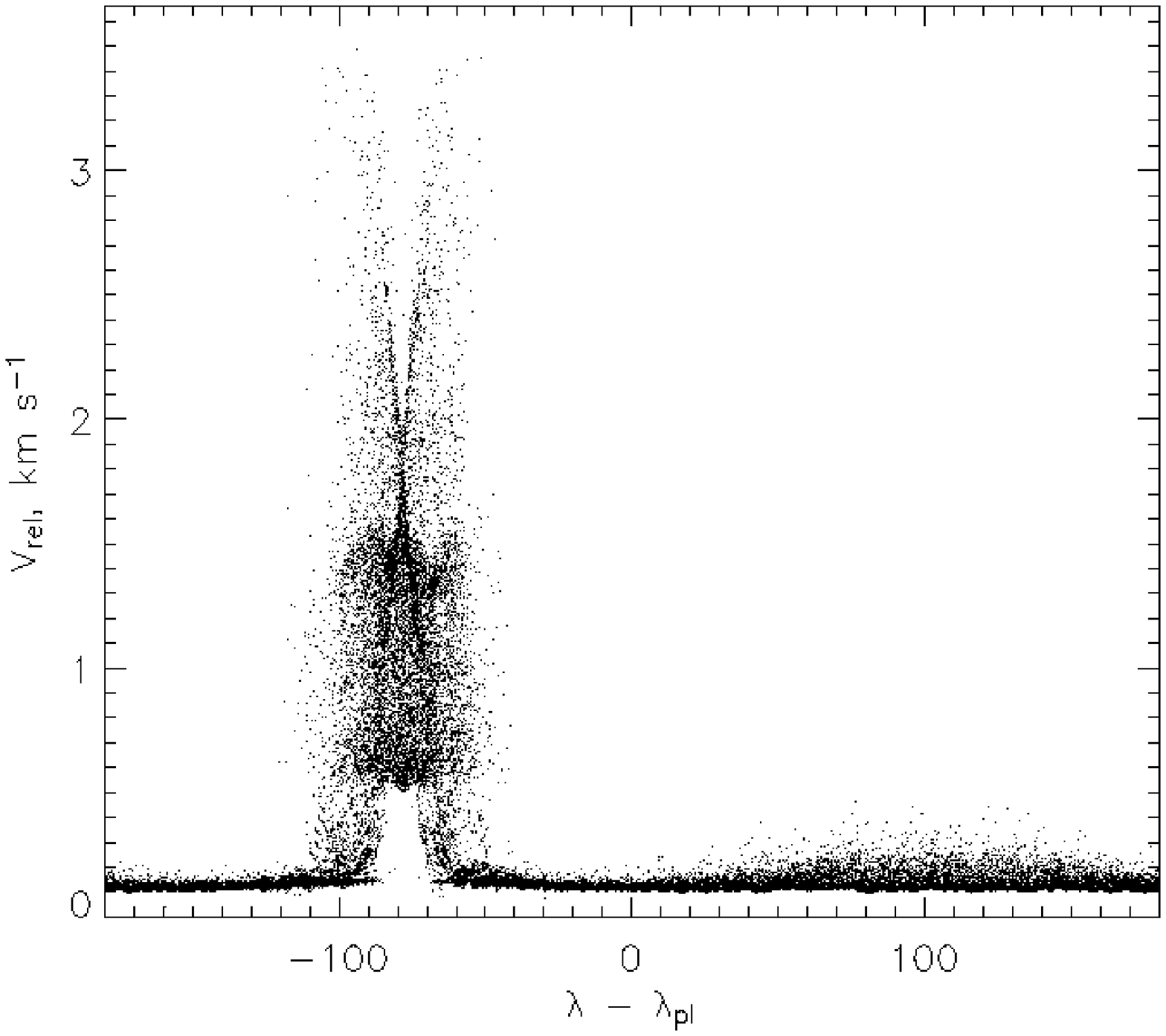} \\[-0.1in]
    \hspace{-0.1in} \includegraphics[width=1.8in]{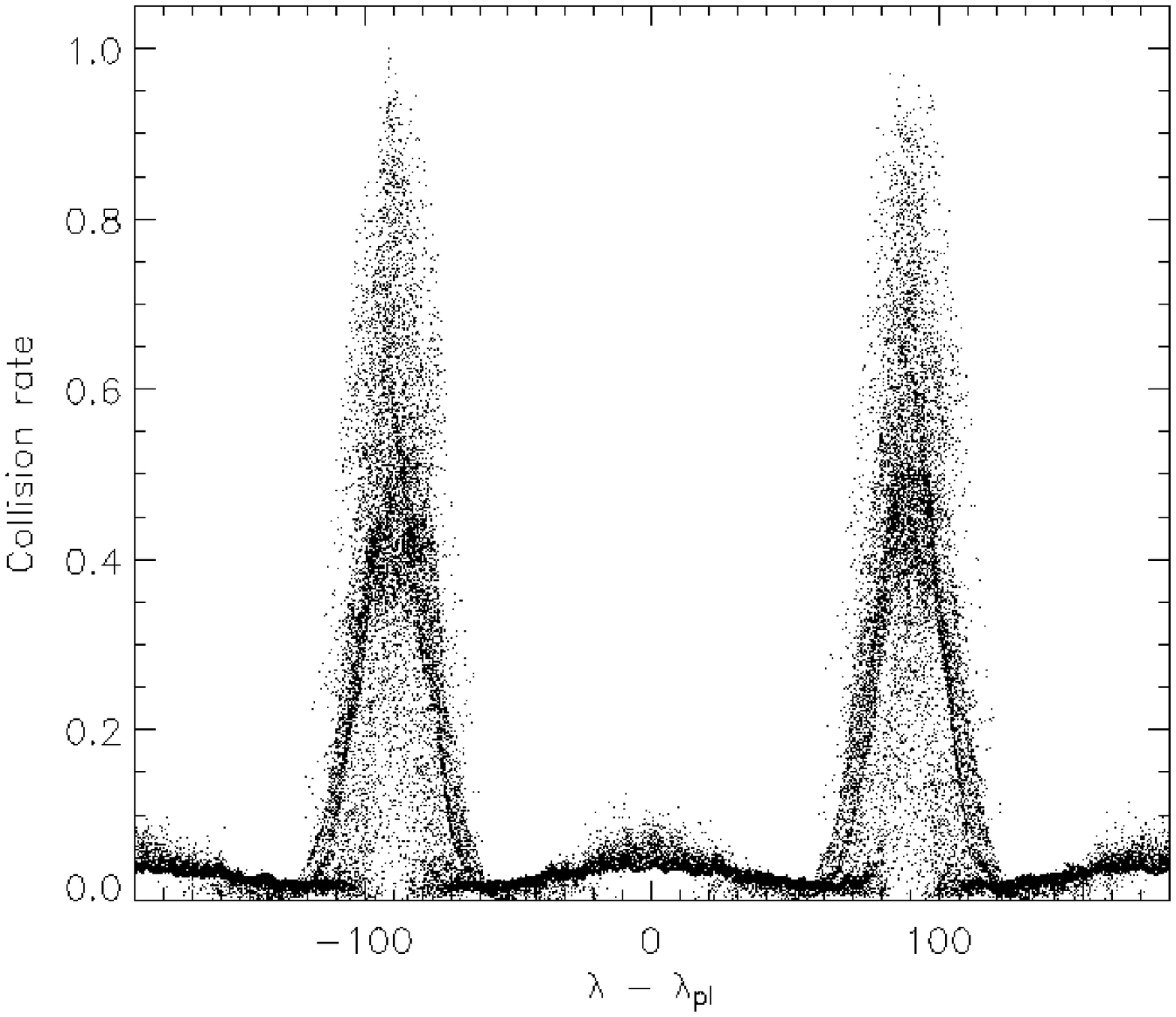} &
    \hspace{-0.2in} \includegraphics[width=1.8in]{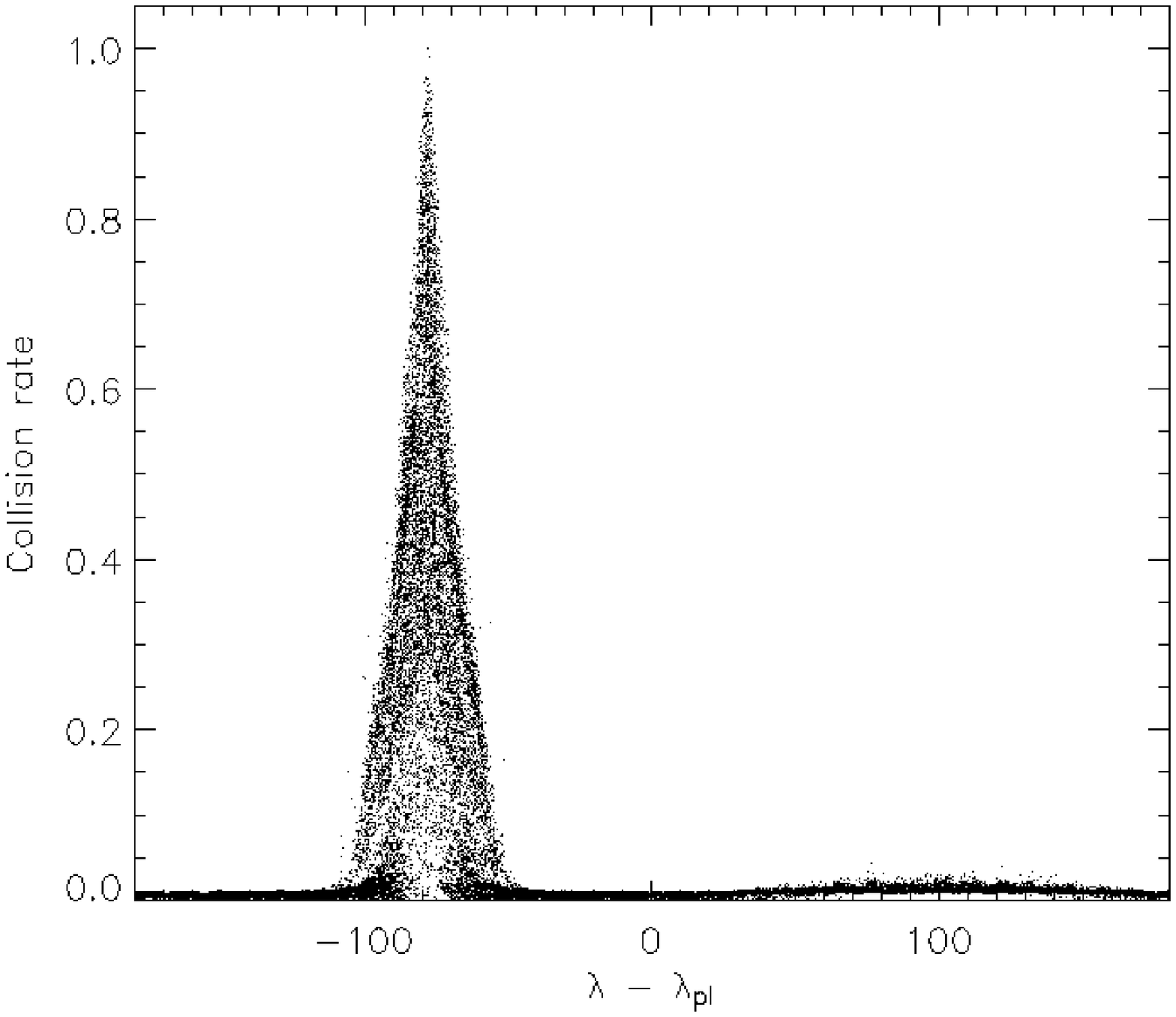} \\
  \end{tabular}
  \caption{Collision rates amongst populations of planetesimals trapped into resonance
  with a $30M_\oplus$ planet that migrated 45-60 AU from a $2.5M_\star$:
  \textbf{(left)} 3:2 resonance;
  \textbf{(right)} 2:1(u) resonance.
  The collision rate (bottom) is the product of the volume density of
  cross-sectional area of planetesimals (top) and the relative velocity
  of collisions (middle).
  Volume densities and collision rates have been normalised to the maximum for all
  planetsimals.
  Each point represents the result for one of 40,000 planetesimals in the disks, and
  is plotted against with the longitude of the planetesimal relative to the longitude of
  the planet.}
  \label{fig:pllcollrates}
\end{figure}

The distribution of grains that are blown out of the
system by radiation pressure as soon as they are created
depends on where those grains are most often created.
In order to quantify the structure of the disk comprised of
grains created in the break-up of large planetesimals, the
following model was devised.
As input, this model took the positions and velocities of a population
of planetesimals that had previously been trapped into resonance by a
migrating planet.
For comparison with the simulations of \S \ref{s:bound},
simulations were considered that placed planetesimals in the 3:2 and
2:1(u) resonances of a $30M_\oplus$ planet which migrated 45-60 AU
from a $2.5M_\odot$ star.
However, rather than using a full integration to determine the outcome
for 200 planetesimals, these populations were simulated using the
approach of W03, since this allowed large numbers of planetesimals
to be simulated without requiring lengthy integrations (40,000 planetesimals
were considered here).
Then for each planetesimal, the rate at which it collides with other
planetesimals was determined using the fact that this collision rate
is proportional to the product of the volume density of cross-sectional
area of nearby planetesimals and the relative velocity of their collisions
(Opik 1951; Wyatt \& Dent 2002).
These two factors were derived by considering the number of planetesimals
within a defined radius (4 AU in this case) of the planetesimal, as well
as the mean of the velocities of those planetesimals relative to the
planetesimal in question.
These factors, as well as the resulting collision rates, are plotted
in Fig.~\ref{fig:pllcollrates}.

Fig.~\ref{fig:pllcollrates} shows that as well as a higher volume density
in the clumps (as expected), there is also a higher relative velocity of
collisions in the clumps.
This means that the collisions are not only more destructive, but also
occur at a higher rate (by more than an order of magnitude) than outside the
clumps.
Thus a planetesimal is most likely to be destroyed at the locations in its orbit
when it is inside one of the clumps (i.e., near pericentre), and
this is where the trajectories of the blow-out grains are most likely to
start.
This means that the distribution of these blow-out grains will not be
axisymmetric.

\begin{figure}
  \centering
  \begin{tabular}{cc}
    \hspace{-0.2in} \includegraphics[width=1.8in]{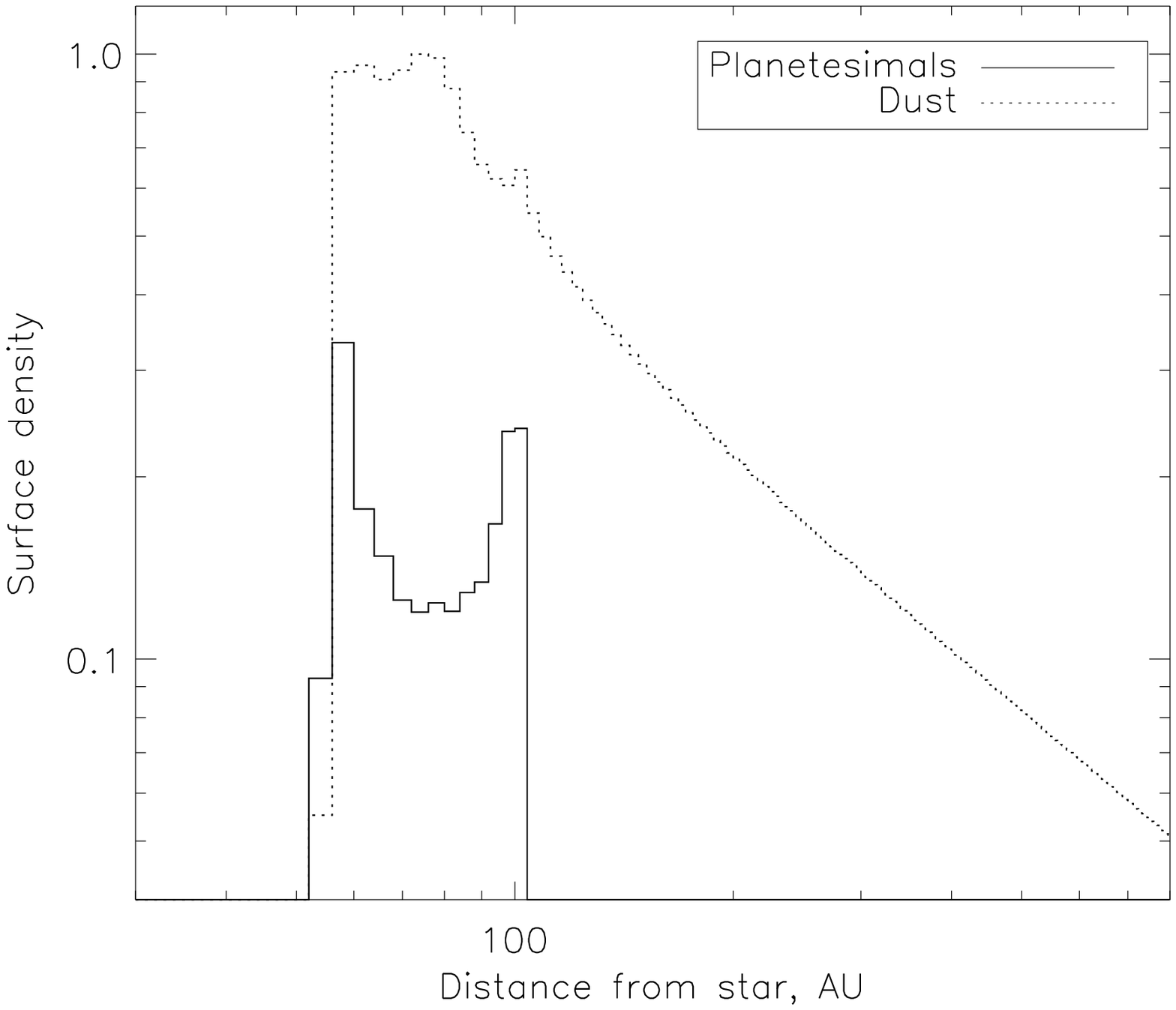} &
    \hspace{-0.2in} \includegraphics[width=1.8in]{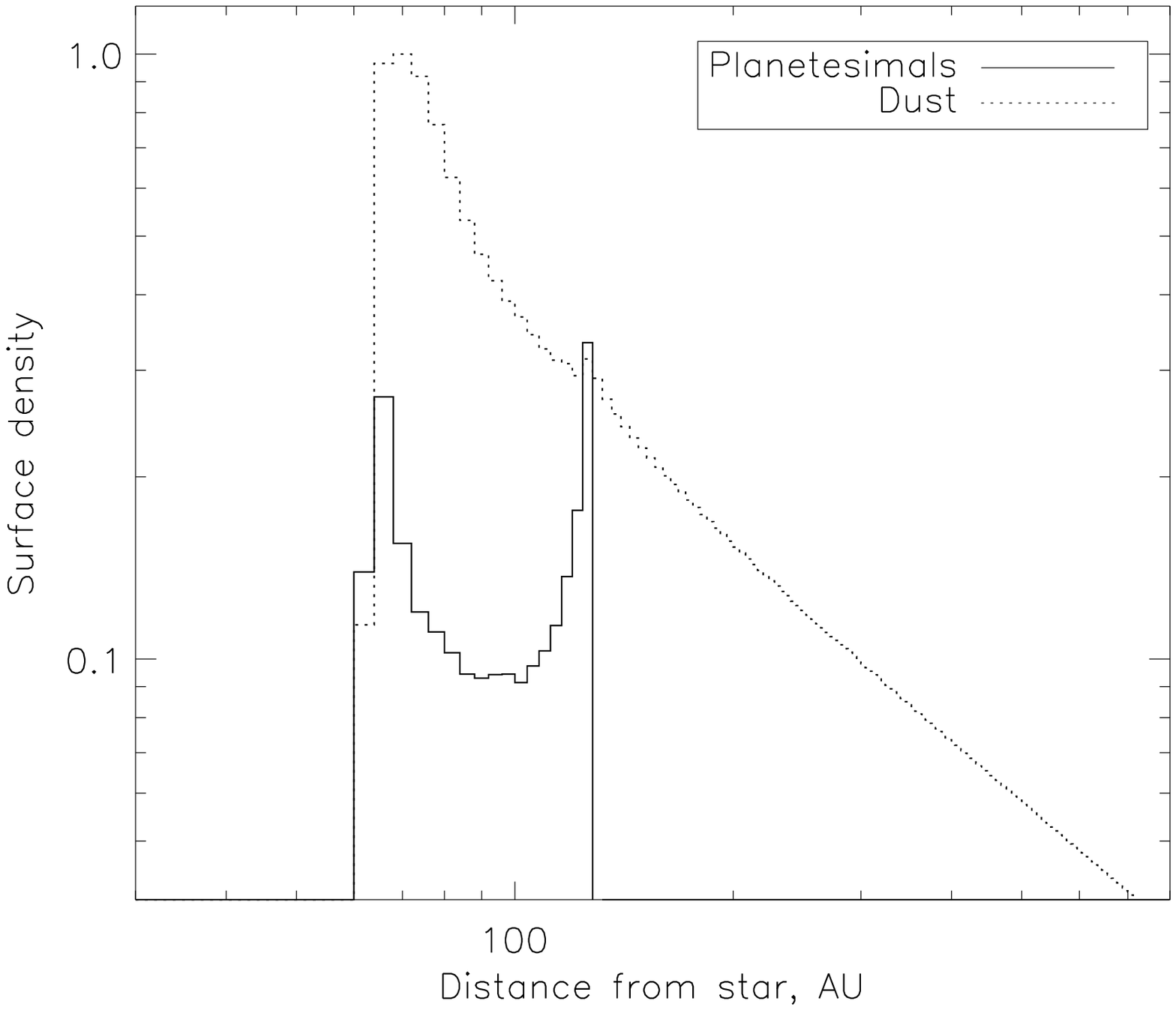} \\[-0.1in]
    \hspace{-0.2in} \includegraphics[width=1.8in]{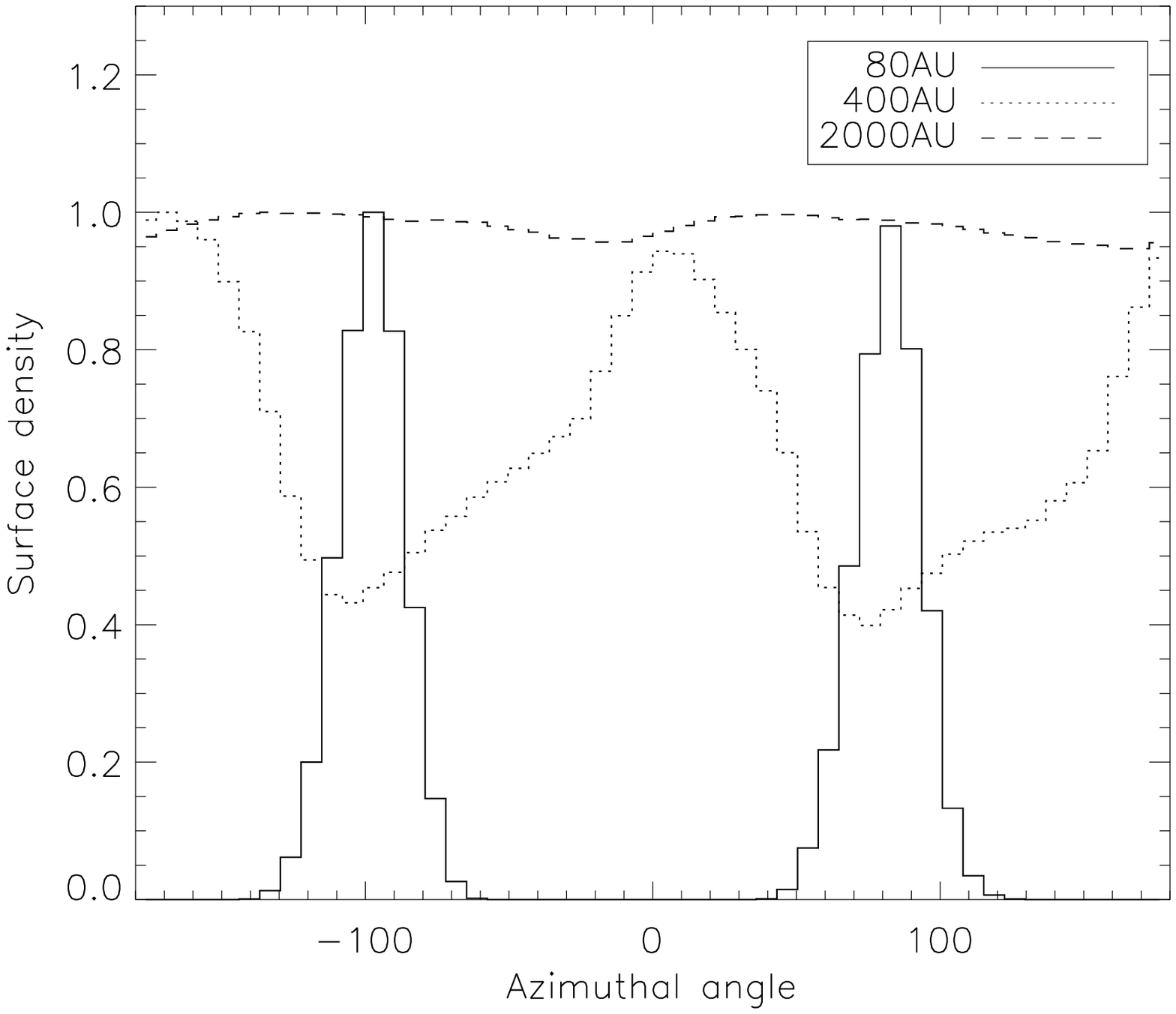} &
    \hspace{-0.2in} \includegraphics[width=1.8in]{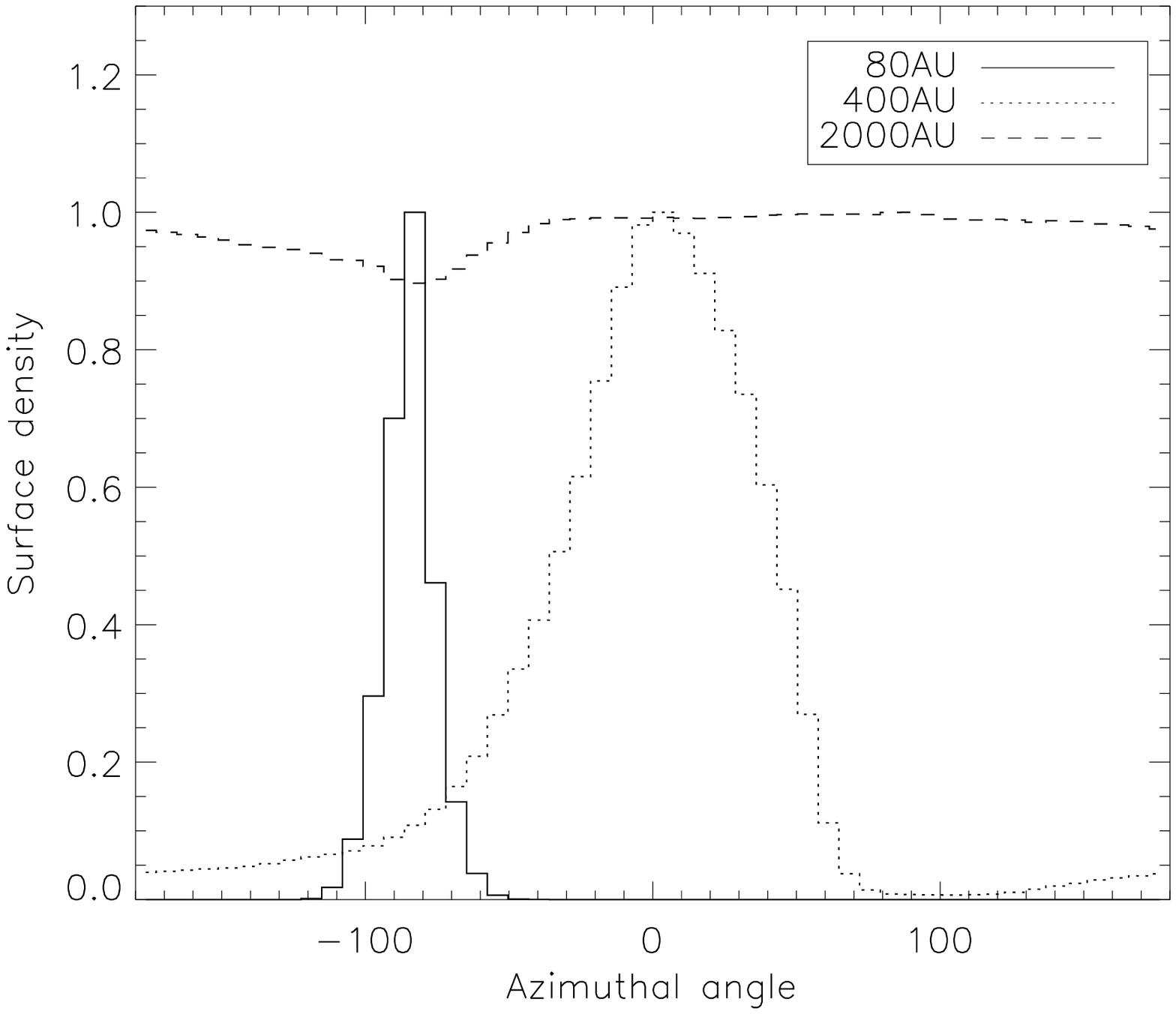} \\
  \end{tabular}
  \caption{Surface density distribution of $\beta=1$ dust grains originating from the
  destruction of planetesimals trapped into resonance
  with a $30M_\oplus$ planet that migrated 45-60 AU from a $2.5M_\star$:
  \textbf{(left)} 3:2 resonance;
  \textbf{(right)} 2:1(u) resonance.
  The radial distribution of surface density is shown in the top plot, and
  the azimuthal distribution at 80, 400 and 2000 AU from the star is shown in
  the bottom plot.
  All distributions are scaled to peak at 1, except the planetesimal distributions
  which peak at 1/3.}
  \label{fig:rphists}
\end{figure}

To determine the spatial distribution of the blow-out grains, the evolution of
the planetesimal population was considered for a total time which
depended how far from the star the dust distribution needed to be considered
(6 planet orbits were considered in this case allowing the distribution to
be determined out to at least 2000 AU).
Since the planetesimal distribution is constant in the frame rotating with the
planet, the positions and velocities of the planetesimal population at subsequent
timesteps were assumed to be the same as those of the planetesimal population at
the end of the migration, but rotated with the planet;
the collision rate of those planetesimals was as determined in Fig.~\ref{fig:pllcollrates}.
In each timestep, a fixed number of dust grains was introduced, with initial
positions and velocities that were the same (suitably rotated) as those of
planetesimals chosen randomly according to the collision rates of the planetesimals.
The evolution of those dust grains was then integrated for the remaining timesteps
under the action of stellar gravity and radiation pressure.
Equilibrium was achieved immediately by assuming the planetesimal that produced
any given dust grain was equally likely to have created one a time equal to the
total integration time ago;
i.e., the integration "starts" with dust grains far from the star.
The result is a movie which shows how the planetesimal and dust population evolves
with time.
Since the dust distribution is constant in the frame rotating with the planet,
this was also coadded in the frame rotating with the planet to determine the spatial
distribution of those grains.
This is shown in the far right plots of Fig.~\ref{fig:spatial} for $\beta=1$ and 10.
It is also shown more quantitatively in Fig.~\ref{fig:rphists} which shows histograms
of the radial and azimuthal distributions of the grains for $\beta=1$.

The radiation pressure blow-out grains form spiral structure which emanates from the
clumps;
i.e., for the 3:2 resonance there are two spirals that start at longitudes $\pm 90^\circ$
from the planet, and the 2:1 resonance exhibits one spiral that starts $\sim 90^\circ$
behind the planet.
Note that in these simulations dust grains were also created when the planetesimals were
outside the clumps, just in much lower quantities.
The spirals unwind in the direction opposite to the planet's motion.
This arises because the pattern speed rotates with the planet (i.e., relatively fast),
and while the dust grains can start with a comparable velocity to the planet's orbital
motion (particularly since they are created near pericentre), their azimuthal motion
is soon much slower than that of the planet as they recede from the star.
Fig.~\ref{fig:rphists} shows how the radial distribution of the $\beta=1$ grains falls
off $\propto r^{-1}$ outside the region where they are being produced, as expected
for grains on hyperbolic orbits;
a similar fall off is seen for $\beta=10$ grains for which the surface density
drops $\propto r^{-1.1}$.
This figure also shows how the spirals becomes less pronounced (i.e., the contrast
between the density in and out of the spiral is lower) at large distances
from the star as the spiral structure diffuses at large distances.
Further it illustrates how the single spiral of the 2:1 resonance is maintained
out to greater distances than that of the 3:2 resonance because it takes longer for
this to merge with the nearest winding.
In fact the spiral structure of blow-out grains will diffuse much faster than
illustrated on Fig.~\ref{fig:rphists}, since the tightness of the winding of the
spiral is determined by the factor $\beta$:
dust grains with higher $\beta$ are accelerated out the system faster than those
with lower $\beta$ meaning that the resulting spiral is less tightly wound.
Representative simulations were performed that made the assumption that the dust grains
produced had $\beta$ chosen randomly from the range 1-10.
Spiral structure close to the clumps is still well pronounced, and only becomes more
diffuse at $\gg 400$ AU.

The distributions shown for $\beta>0.5$ in Fig.~\ref{fig:spatial} are only valid for
a disk comprised only of the planetesimals considered in that simulation (i.e.,
all trapped in the same resonance with a narrow range of eccentricities).
This is because the structure is determined by the collision rates of those
planetesimals.
A real disk is not just comprised of planetesimals in one resonance, and
the interaction of the resonant plantesimals with planetesimals that are either
non-resonant, or in other resonances, will affect the collision rates.
To assess this, we also performed a representative simulation of the distribution
of dust grains expected as a result of the planetary migration described in W03
to explain the structure of the sub-mm emission from the Vega disk.
In this simulation planetesimals were trapped into a variety of resonances (though
mainly the 3:2 and 2:1(u)), and many remained on non-resonant orbits.
The order of magnitude increase in collision rate is still apparent in the clumps,
and so two armed spiral structure dominates the distribution of blow-out grains.
Since the two clumps are of different magnitude it is also worth noting that one
of the spirals is more dense by a factor greater than simply the ratio of the 
clump densities, since the production rate of those particles is proportional
to the product of the clump density squared and the relative velocity of the
collisions (which is slightly higher in the brighter clump because of the
superposition of the 3:2 and 2:1(u) resonances there).

The model discussed so far in this section is only valid for the $\beta>0.5$
dust grains that are created by the destruction of large planetesimals.
Such dust grains could also be created in the destruction of intermediate
sized dust grains that are still on bound orbits.
As Fig.~\ref{fig:spatial} showed, those intermediate sized grains could have
a distribution that is axisymmetric.
The blow-out grains arising from the intermediate sized grains, with
$\beta>\beta_{crit}$, would be expected to have a radial distribution similar
to that shown in Fig.~\ref{fig:rphists}, but with an axisymmetric azimuthal
distribution.

\section{Discussion}
\label{s:disc}

\subsection{Summary of Grain Populations}
\label{ss:popsum}
In \S \S \ref{s:bound} and \ref{s:unbound} the distribution
of small dust grains resulting from the destruction of planetesimals
that are trapped in resonance with a planet were studied showing
how they differ significantly from that of the planetesimals
themselves.
Here the results are summarised.
There are three distinct grain populations each of which exhibiting a different
spatial distribution:

\begin{itemize}
  \item
\textbf{(I)} Large grains with $\beta>\beta_{crit}$ (i.e., typically $>$ a few mm)
have the same clumpy resonant distribution as the planetesimals, albeit one
which is slightly azimuthally smeared out for $\beta=(0.2-1)\beta_{crit}$;
  \item
\textbf{(II)} Moderate sized grains with $\beta_{crit} < \beta < 0.5$
(i.e., typically a few $\mu$m to a few mm) are no longer in resonance and
have an axisymmetric distribution which is also more radially extended and
vertically broadened than that of population I grains.
Such grains may also have a short lifetime due to the increased chance of a
close encounter with the planet;
  \item
\textbf{(III)} Small grains with $\beta>0.5$ (i.e., typically $<$ a few $\mu$m)
are blown out of the system by radiation pressure immediately on creation and so
have a density distribution which falls off as $\tau \propto r^{-1}$, however
the structure of these grains can be further divided into two subclasses:
\textbf{(IIIa)} grains produced in the destruction of population I grains that
exhibit trailing spiral structure which emanates from the resonant clumps,
and \textbf{(IIIb)} grains produced from population II grains that have an
axisymmetric distribution.
\end{itemize}

\subsection{Predictions for Multi-waveband Imaging}
\label{ss:multiwav}
Observations in different wavebands are sensitive to different
sizes of dust grains.
Thus a disk could exhibit different structures when imaged in
different wavebands, providing the two wavelengths sample different
grain populations.
This has the potential to provide a valuable observational test of models
which explain clumpy structure seen in sub-mm observations of debris
disks as due to resonant trapping of planetesimals.
It is also relevant to ask whether the sub-mm observations would sample
dust grains with a spatial distribution that is similar to that of their
parent planetesimals (i.e., population I grains), as was
assumed in W03.

The size of grains sampled in different wavebands depends to some
extent on the composition of the grains, but is most strongly dependent
on the size distribution of grains in the disk.
At present this size distribution is not predicted by the models.
The most simple assumption is that, since the dust is produced
in the collisional destruction of planetsimals, the size distribution
is the same as that expected in an infinite collisional cascade with
$n(D) \propto D^{-3.5}$ down to the radiation pressure blow-out limit
(Wyatt \& Dent 2002).
Although in practise it has been shown that an abrupt cut-off to the size
distribution at the blow-out limit would cause a wave in this distribution
(Th\'{e}bault, Augereau \& Beust 2003).
This simple assumption can be used to explore the relative importance of grains
in populations I and II, but does not make any
predictions about the quantity or observability of population III grains.
It also neglects the fact that the dynamical lifetimes of population
II grains may be lower than their collisional lifetimes due to
the increased chance of scattering by the planet, and their collisional
lifetimes may be affected by the broadening of their spatial distribution.

However, this simple size distribution has been shown to provide a good
fit to the spectral energy distributions of the emission from several
debris disks (Wyatt \& Dent 2002, Sheret, Dent \& Wyatt 2004).
The results presented in Fig.~5 of Wyatt \& Dent (2002) can also be used
to work out which grain populations we would be seeing in the Fomalhaut
disk for different wavebands (assuming of course that some of the planetesimals
in this disk are trapped in resonance with a planet, which has not been
proved).
Since just 5\% of the sub-mm emission comes from grains either
$<300$ $\mu$m or $>20$ cm in diameter, with roughly equal weightings in
this range toward different bins of log particle diameter, this implies
that sub-mm observations would indeed be dominated by the population
I grains that have a similar spatial distribution to that of
the planetesimal population.
However, this also implies that, since less than 5\% of the
25-100 $\mu$m emission comes from grains that are bigger than
$\sim 6$ mm with weightings that favor the lower end of the size range,
such observations would be dominated by population
II grains and would have an axisymmetric spatial
distribution.

Thus for disks where this size distribution holds, and in which clumps are
seen in sub-mm images, a model which interpreted the clumps as the result
of planet migration and the consequent trapping of planetesimals into the
planet's resonances could be tested by seeing if mid- to far-IR imaging
of the disk shows it to be axisymmetric.
By consideration of equation (\ref{eq:dres}) it is possible to infer
that the wavelength at which the transition from clumpy to smooth structure
occurs is indicative of both the mass of the perturbing planet, with the
transition shifted to shorter wavelengths for more massive planets,
and the spectral type of the star, with shorter wavelength transitions
for lower mass stars.
In other words multi-wavelength imaging could also reveal the mass of the
planet.
In fact Fomalhaut's disk appears asymmetric at a range of wavelengths
(Holland et al. 2003; Stapelfeldt et al. 2004; Marsh et al. 2005;
Kalas et al. 2005), favoring a mechanism that affects grains of all
sizes such as an offset centre of symmetry due to the non-circularity
of a perturbing planet's orbit (Wyatt et al. 1999; Marsh et al. 2005;
Kalas et al. 2005).

To assess the importance of the population III grains relative
to that of the populations I and II grains described
above, a more detailed model of the size distribution of the disk would be
required which is beyond the scope of this paper, although some considerations
are discussed in \S \ref{ss:vega}.

\subsection{Implications for Vega}
\label{ss:vega}
Vega's dust disk has been observed to be clumpy when imaged in the
sub-mm (Holland et al. 1998), yet more recent observations show
its structure to be axisymmetric at mid- to far-IR wavelengths
(Su et al. 2005).
This does not in itself validate the model of W03, since
Su et al. also showed that the disk is much more extended at these
wavelengths than in the sub-mm.
They devised a three component model to explain the radial emission
distribution observed from 25-850 $\mu$m which is comprised of
grains that are 4, 36 and 430 $\mu$m in diameter.
The two smallest sizes have an optical depth distribution that falls off
$\tau \propto r^{-1}$ and are small enough that $\beta>0.2$.
Such grains dominate the mid- to far-IR images and are interpreted as
grains that are in the process of radiation pressure blow-out (i.e.,
population III grains), while the larger grain population, with a
density distribution peaked around 100 AU is required to fit the 850
$\mu$m images (and must be predominantly population I grains to
produce the observed clumpy structure).
Su et al. interpreted this as evidence of a recent collision having
occured in the disk, since the blow-out grains are short-lived and it is
unfeasible for the observed outflow to have been continually replenished
over Vega's 350 Myr lifetime.
However, they did not attempt to explain the azimuthal structure in
the sub-mm images, or lack thereof in the short wavelength structure.
Here the model of Su et al. is expanded to turn the 3 component
model into one with a size distribution extending across all sizes.
The aim is to provide a more realistic description of the disk
which can be used to question whether, in the absence of a model
that predicts the size distribution in the W03 model, the relative
quantities of populations I, II and III grains are physically plausible
in the context of that model given our understanding of size distributions in
collisional cascades, or if another model needs to be sought to explain
the clumpy sub-mm structure.

The size distribution in Vega's disk is assumed to be defined by different
power laws in three size ranges considered to represent populations
I, II and III grains.
Su et al. already showed that the distribution of population III
grains follows a size distribution of $n(D) \propto D^{-3.0}$ in
the range $\sim 4-50$ $\mu$m and to have a distribution $\tau \propto r^{-1}$
for $r>86$ AU.
This distribution is maintained here, but the minimum and maximum grain
sizes, as well as the total cross-sectional area, are left as variables.
The population I and II grains are fixed to lie at 100 AU, the peak
in the optical depth distribution found by W03, and the 
size distributions of both populations are assumed to follow that of a
collisional cascade with $n(D) \propto D^{-3.5}$, but with different
amounts of cross-sectional area in each of the populations.
The division between populations I and II was set at 1 mm, a
changeover size consistent with a planet of size $40-50M_\oplus$
(eq.~\ref{eq:dres}).
Following Su et al., astronomical silicates were assumed for the particles'
optical properties.
Since Su et al. already showed that this fits the surface brightness
distribution, it remains to fit the spectral energy distribution (SED),
where the emission from population III grains is taken to be that out to
1000 AU.
An additional constraint was set to determine the relative contributions
of the population I and II grains, which is that both of these populations
combined contribute $\sim 50$\% of the total flux at 160 $\mu$m as found
in the modeling of Su et al.

\begin{figure}
  \centering
  \begin{tabular}{rlrl}
    \hspace{-0.1in} \textbf{(a)} &
    \hspace{-0.5in} \includegraphics[width=1.8in]{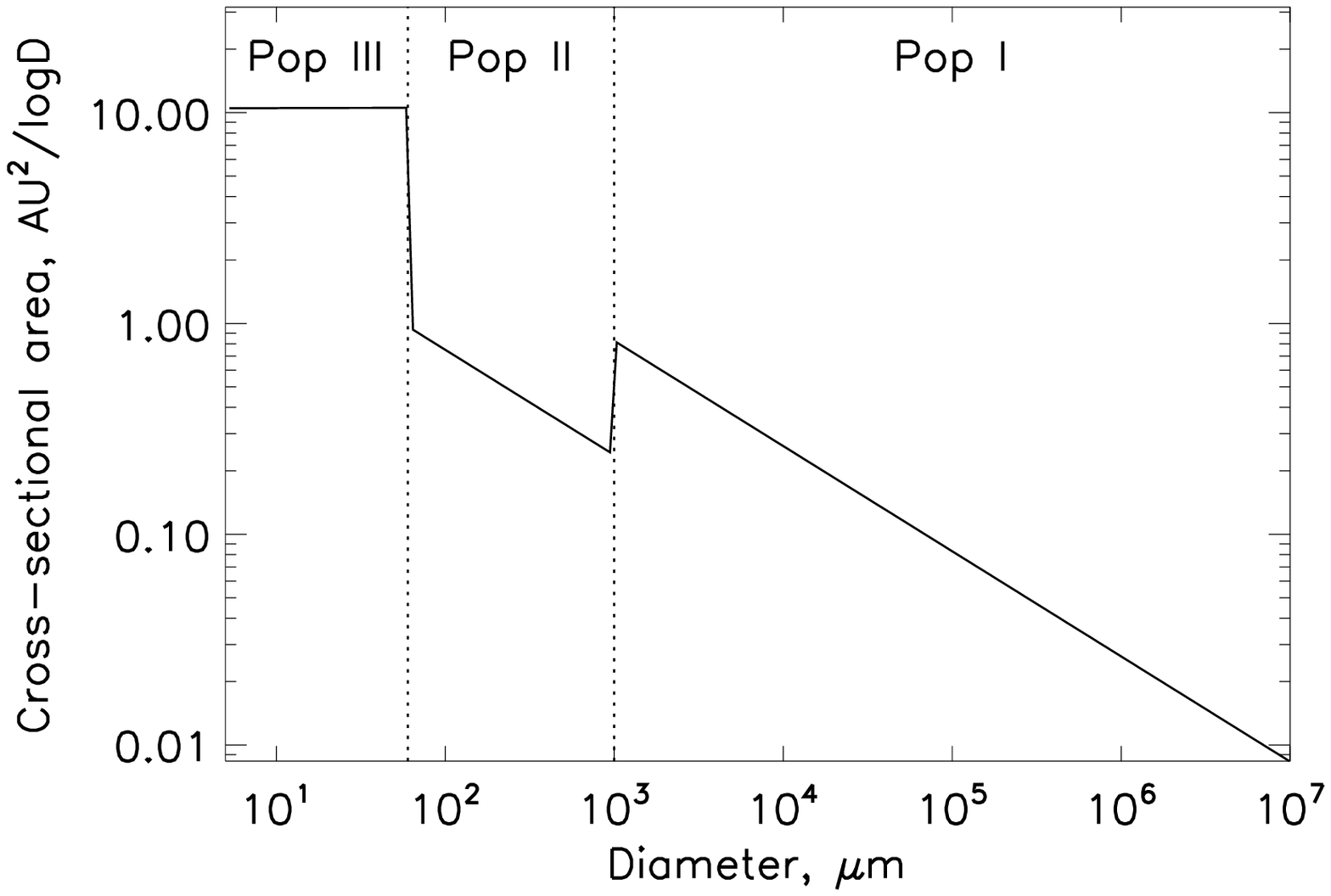} &
    \hspace{-0.1in} \textbf{(c)} &
    \hspace{-0.2in} \includegraphics[width=1.8in]{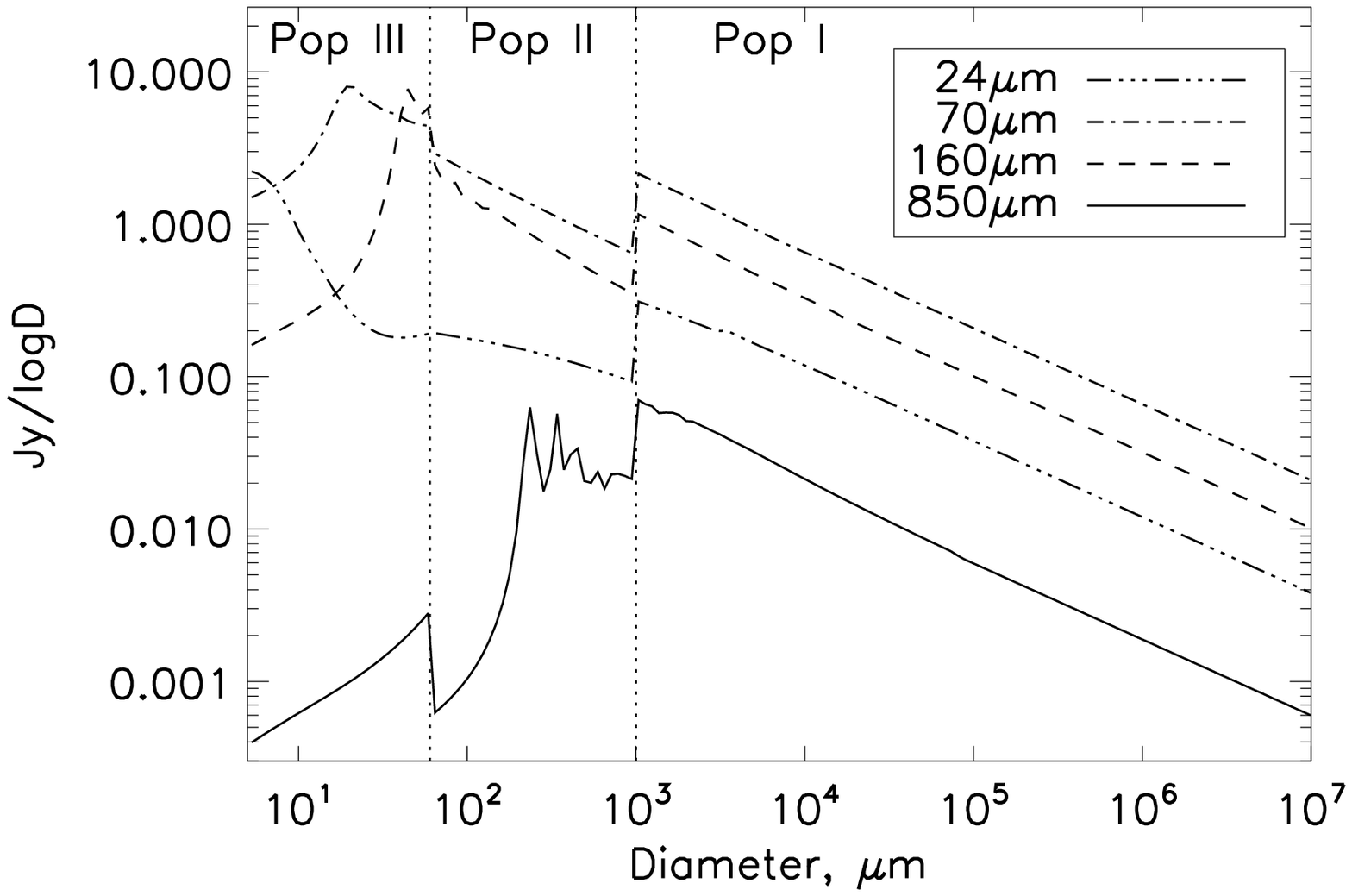} \\[-0.0in]
    \hspace{-0.1in} \textbf{(b)} &
    \hspace{-0.5in} \includegraphics[width=1.8in]{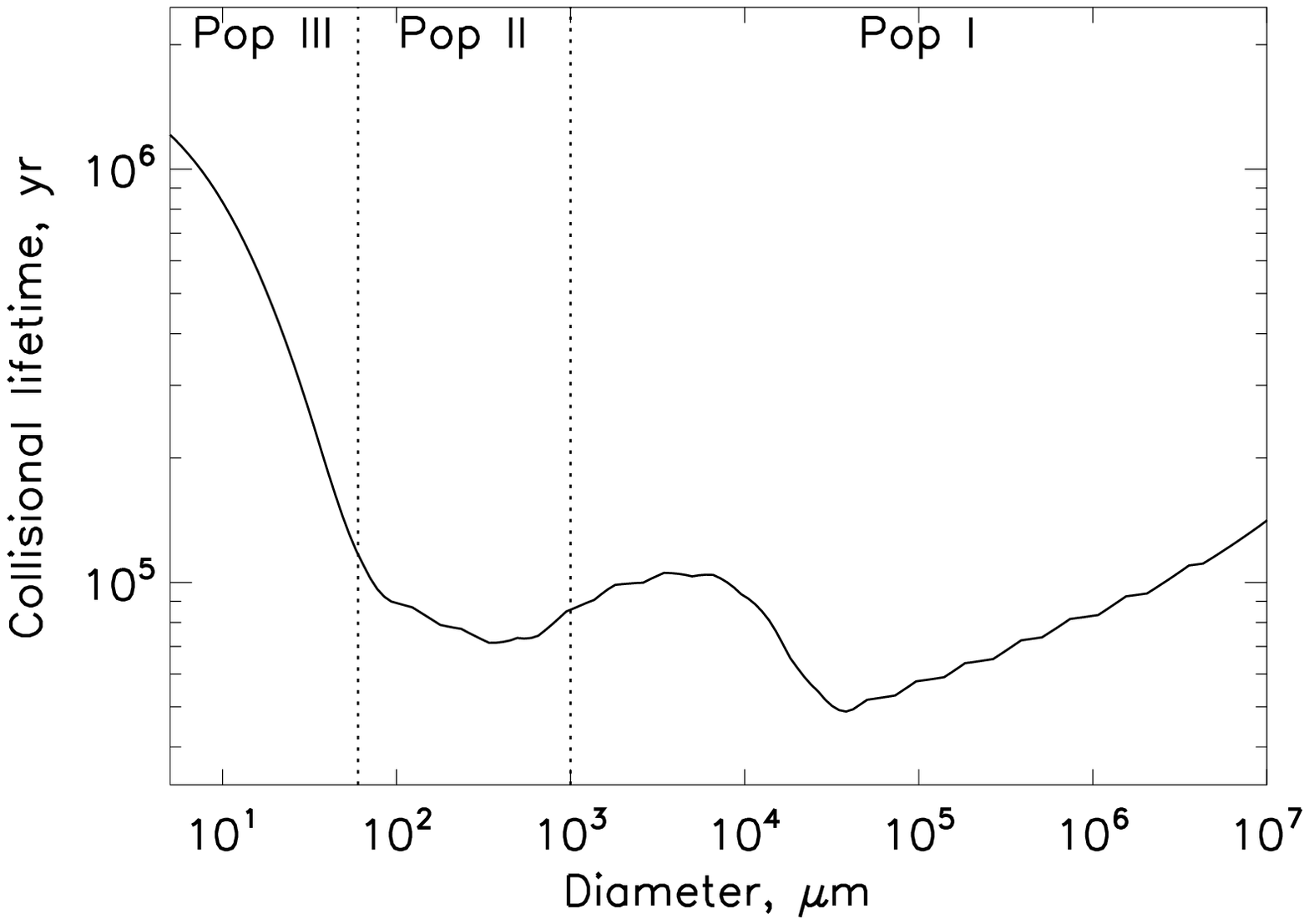} &
    \hspace{-0.1in} \textbf{(d)} &
    \hspace{-0.2in} \includegraphics[width=1.8in]{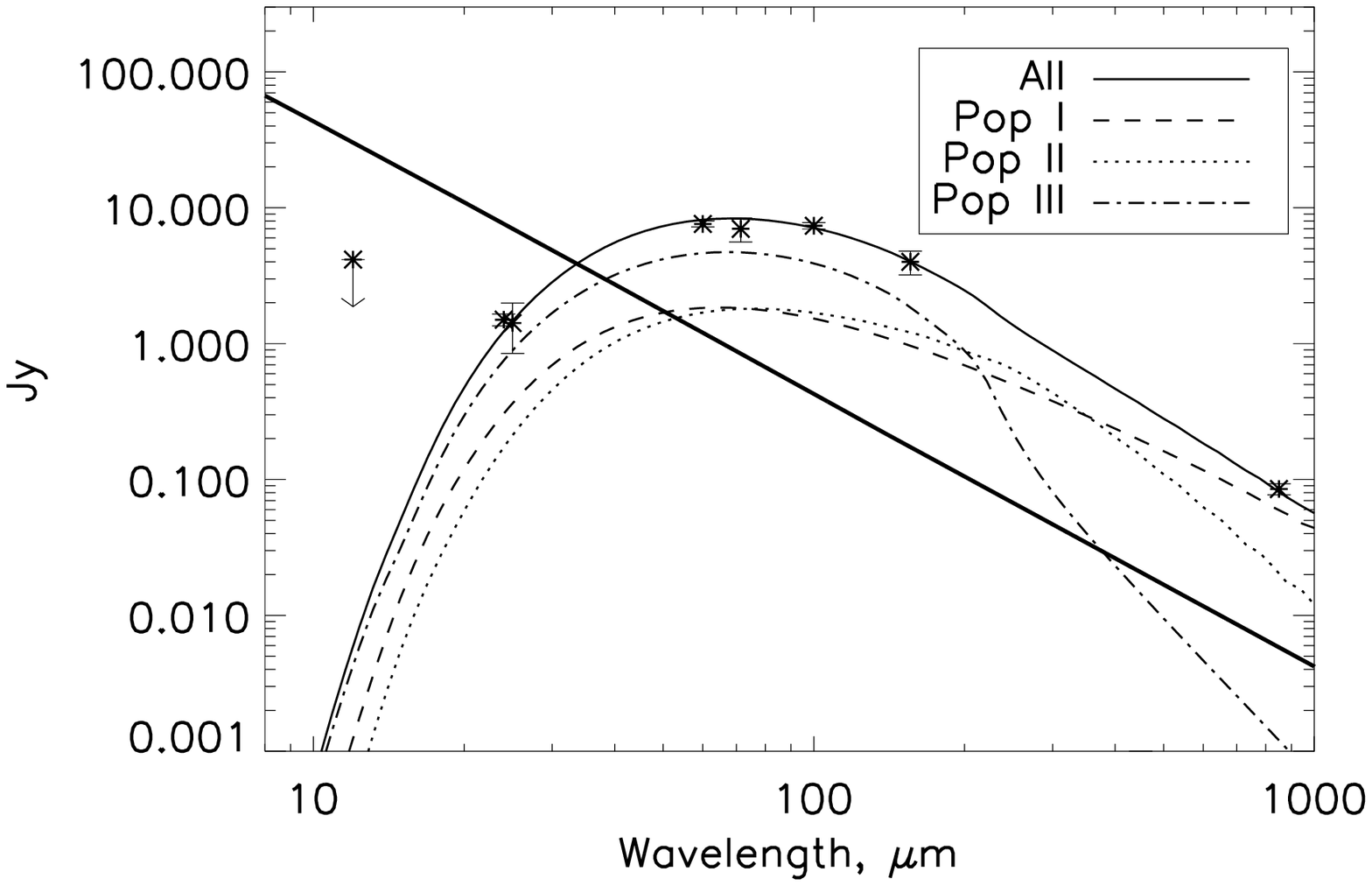} 
  \end{tabular}
  \caption{Model for the size distribution of dust in the Vega disk:
  \textbf{(a)} Cross-sectional area as a function of particle size, with
  boundaries indicated between the different populations I, II and III.
  For population III the area is the total out to 1000 AU;
  \textbf{(b)} Collisional lifetime of different sized grains in the
  disk, assuming the collisional properties of weak ice from Wyatt \& Dent
  (2002), and an average collision velocity of 1.4 km/s and a conversion
  of cross-sectional area to volume density that assumes this area is
  spread evenly around a torus extending 86-200 AU from the star for
  populations I and II, and only considering the area of population III
  grains in the same region; 
  \textbf{(c)} Contribution of different grain sizes to the total flux
  in the different Spitzer and SCUBA wavebands;
  \textbf{(d)} Spectral energy distribution of emission from the Vega
  disk model (solid line) showing the contribution of different populations
  to that spectrum (dotted, dashed and dash-dot lines).
  Fluxes are excess fluxes after photospheric subtraction from the IRAS
  Faint Source Catalog, Spitzer (Su et al. 2005), and SCUBA (Holland et
  al. 2005).
  The level of the photospheric emission is shown with the thick solid line.}
  \label{fig:sizedist}
\end{figure}

The size distribution which fits the observed emission spectrum
(and by analogy with the modeling of Su et al., also the surface
brightness distributions), is shown in Fig.~\ref{fig:sizedist}a.
The contribution of different size grains in the distribution to
the fluxes in the Spitzer and SCUBA wavebands is shown in
Fig.~\ref{fig:sizedist}c, and the modeled emission spectrum, showing
the contribution of the different populations, is shown in
Fig.~\ref{fig:sizedist}d.
Because of the way the emission efficiency falls off for population
II grains at longer wavelengths, the condition of equal contributions
from bound and unbound grains to the 160 $\mu$m flux (and the need to
fit the 850 $\mu$m flux) results in a size distribution with a
level of population II grains some 3-4 times lower than
that expected in an idealised collisional cascade (e.g., that assumed
in Wyatt \& Dent 2002).
This also results in the population II grains contributing just
$\sim 25$\% of the 850 $\mu$m flux, in line with the observation
that this image is clumpy (Holland et al. 1998).
Increasing the relative amount of population II grains both increases
their contribution to the 850 $\mu$m image and of bound grains to the
total 160 $\mu$m flux.
It is this result, along with the quantity of material
in population III relative to populations I and II, that needs to be
explained within the context of collisional cascades.
The results for the boundaries defining population III grains are
less important, since these values depend on the optical properties
of the grains which are not considered here.
\footnote{The change-over from population II to III in this
model occurs at 60 $\mu$m, which is at $\beta = 0.17$.
A cut-off at such a low $\beta$ may be real, since a cut-off at
$\beta<0.5$ is expected if, as shown in Fig.~\ref{fig:pllcollrates},
most grains are created at the pericentre of highly eccentric orbits
(since such grains are put onto hyperbolic orbits provided
$\beta > (1-e)/2$).
However, this small discrepancy could equally be removed by changing
the optical properties used in this model.}

To answer whether the distribution of Fig.~\ref{fig:sizedist}a is
a realistic, or even expected, size distribution in Vega's collisional
cascade, Fig.~\ref{fig:sizedist}b shows the lifetimes of the different
grains due to collisions.
This plot exhibits some well known features:
starting at the largest sizes, the collisional lifetime of grains
is reduced as smaller sizes are approached due to the greater quantities of
cross-sectional area of smaller objects in a collisional cascade.
The lower numbers of population II grains, however, cause an increase
in the lifetimes of the population I grains that should have been
destroyed by this population (Th\'{e}bault et al. 2003).
The population II grains have a lower lifetime than the population I
grains because these are destroyed in collisions with the large quantities
of blow-out grains in this distribution (Krivov, Mann \& Krivova 2000).
However, the lifetime of the population III grains themselves due to
collisions is much lower than their blow-out time indicating that only
a small fraction of the population III grains are themselves destroyed
in collisions on their way out.
This means that it may be possible to explain the break in the size
distribution between populations I and II, as inferred from the SED
modeling, by the destruction of population II grains in collisions
with those in the process of blow-out by radiation pressure
(Krivov et al. 2000), and the knock-on effect induced by the lower levels
of such grains in the distribution at larger sizes (Th\'{e}bault et al.
2003).
This effect may be further accentuated by a lifetime for population II grains
that is even shorter than their collisional lifetimes of $\sim$ 100,000 years
which may be caused by interaction with the planet which occurs on
1000 year timescales (eq.~\ref{eq:tscat}).

However, this still leaves the question of why there are so many
population III grains.
This modeling finds a similar mass loss rate to Su et al. for this
population of $\sim 2M_\oplus$/Myr, which, as Su et al. point out,
implies that the cascade must have been initiated relatively recently.
However, the lack of evidence for non-axisymmetry in the far-IR emission
at large distances from the star seen by Su et al. poses problems for
the W03 model.
This is because the mass of population II grains is comparable to that
of population III grains at $\sim 2 \times 10^{-3}M_\oplus$, yet the
population II grains are only destroyed in collisions on timescales of
$\sim$ 100,000 years whereas population III grains
are removed on 1000 year timescales.
This means that, unless there is some mechanism that turns population II
grains into population III grains on 1000 year timescales, then at most
1\% of the population III grains can be of type IIIb.
The remainder must be population IIIa grains, and the collision
rates indicate that this is not unreasonable:
for the distribution assumed in Fig.~\ref{fig:sizedist}, the mass
in population I grains ($0.6M_\oplus$) is processed in collisions at a
rate $\sim 7M_\oplus$/Myr.
While most of this mass is likely redistributed within population I
rather than lost to population III, it is noted that this rate
would be much larger if the distribution had been assumed to extend
to sizes larger than 10m.
Thus it is possible that the required loss rate could be achieved
with just a few \% of the mass of population I objects being put into
blow-out grains in destructive collisions as long as the population I
distribution extends to large enough objects.
This means that the mid- to far-IR emission should exhibit spiral
structure rooted in the clumps seen in the sub-mm.
Limits on the circularity of the emission detected by Spitzer were
not discussed in detail in Su et al., but it is possible that these
images were not of sufficiently high resolution and/or sensitivity (or
are too confused by the point-like photospheric emission) to rule out
the presence of population IIIa grains.
If this is the case, then the model predicts that at high
resolutions the disk should exhibit spiral structure when
imaged at far-IR and mid-IR wavebands, e.g., when imaged at
25 $\mu$m using a coronagraph with MIRI on the JWST (Wright et al. 2003).
A more detailed confrontation of the model with the Spitzer and
SCUBA observations of the Vega disk is left for a future paper
(Wyatt, Su, Rieke, Holland, in prep.).

\section{Conclusions}
\label{s:conc}
This paper shows how the distribution of small dust grains resulting from
the destruction of planetesimals that are trapped in resonance with a
planet differ from that of the planetesimals themselves, both in terms
of their orbital characteristics and consequently their spatial
distributions.
Three different grain populations are identified based on
grain size:
population I grains that are large enough to remain in the resonance
of the parent object (and so have a clumpy distribution);
population II grains that, due to radiation pressure, are still on
bound orbits, but are no longer in resonance (and so have an axisymmetric
distribution);
and population III grains that are removed from the system by
radiation pressure on short timescales (and so have a distribution
that falls off $\propto r^{-1}$).
Subclasses are defined for population III grains based on the population
designation of the parent object:
population IIIa grains originate in the destruction of population I
grains and exhibit trailing spiral structure emanating from the clumps;
population IIIb grains originate in the destruction of population II
grains and have an axisymmetric distribution.

The fact that a planetesimal belt is made up of particles from
all populations and subclasses, each of which has its own (quite
different) dynamical and spatial distribution implies that observations
in different wavebands can be dominated by different populations and
so exhibit different morphologies.
Adoption of a simple collisional cascade size distribution with no blow-out
(population III) grains implies that if some planetesimals in the disk are
trapped in resonance with a planet then sub-mm observations would trace the
distribution of those planetesimals through population I grains (validating the
approach of Wyatt 2003 in modeling the SCUBA observations of Vega), but that
mid- to far-IR observations would trace population II grains and so have an
axisymmetric distribution.
The wavelength at which the transition occurs from clumpy to smooth
structure is indicative of the mass of the planet.

The size distribution of Vega is modeled in the light of recent Spitzer
observations which show that significant quantities of population III grains
are present (Su et al. 2005).
This shows that there is a significant lack of population II
grains in this distribution (a factor 3-4 under that expected in a
collisional cascade).
Analysis of the collisional lifetimes indicates that this may be due to the
destruction of these grains by those that are in the process of being blown
out by radiation pressure.
It is argued that unless there is some mechanism that is destroying
the population II grains on 1000 year timescales, then the population
III grains, and so the mid- to far-IR images of the Vega disk, should
exhibit spiral structure emanating from the clumps seen in the sub-mm
images.
Detection of such structure may be possible with Spitzer or with MIRI on
the JWST, and would confirm the interpretation
of the morphology of Vega's disk in terms of planetesimals trapped in
resonance with a planet orbiting at 65 AU (Wyatt 2003) as well as indicate
the direction of its motion.

Multi-wavelength imaging thus provides a method of confirming models
interpreting clumps in debris disks as indicative of planetesimals
trapped in resonance with an unseen planet.
Such images can also provide information on the mass and direction
of motion of the perturbing planet, and do not require the
decade timespans of multi-epoch methods for confirming these models
by checking for orbital motion of the clumps (e.g., Ozernoy et al.
2000).




\end{document}